# Ocean Wind Wave Climate Responses to Wintertime North Atlantic Atmospheric Transient Eddies and Low Frequency Flow

M. Y. Markina[1,2, *], J. H. P. Studholme[1], and S. K. Gulev[1,2]

1. Shirshov Institute of Oceanology, Russian Academy of Sciences
2. Lomonosov Moscow State University

*Moscow, Russian Federation*

* Corresponding author address: Margarita Markina, Shirshov Institute of Oceanology, RAS. 36 Nakhimovsky ave. 117997. Moscow. Russian Federation.

Email: markina@sail.msk.ru

**ABSTRACT**

Atmospheric transient eddies and low-frequency flow contribution to the ocean surface wave climate in the North Atlantic during boreal winter is investigated (1980 - 2016). We conduct a set of numerical simulations with a state-of-the-art spectral wave model Wavewatch III forced by decomposed wind fields derived from the ERA-Interim reanalysis (0.7° horizontal resolution). Synoptic-scale processes (2-10 day bandpassed winds) are found to have the largest impact on the formation of wind waves in the western mid-latitude North Atlantic along the North American and western Greenland coasts. The eastern North Atlantic is found to be influenced by the combination of low-frequency forcing (>10 day bandpassed winds) contributing up to 60% and synoptic processes contributing up to 30% to mean wave heights. Mid-latitude storm track variability is found to have a direct relationship with wave height variability on the eastern and western margins of the North Atlantic in particular implying an association between cyclone formation over the North American Eastern Seaboard and wave heights anomalies in the eastern North Atlantic. A shift in wave height regimes defined using an EOF analysis is reflected in the occurrence anomalies in their distribution.

Results highlight the dominant role of transient eddies on the ocean surface wave climatology in the mid-latitude eastern North Atlantic both locally and through association with cyclone formation in the western part of the basin. These conclusions are presented and discussed particularly within the context of long-term storm-track shifts projected as a possible response to climate warming over the coming century.

# 1. Introduction

Ocean surface wind waves are an important component of the climate system, influencing upper-ocean turbulence and mixing (Babanin, 2006, 2009), heat and momentum air-sea fluxes (Veron, 2008, 2011; Sullivan and McWilliams, 2002), the production of atmospheric aerosols via bubbles and sea-spray (de Leeuw et al. 2011; Babanin, 2011), sea ice formation and melting (Fan et al., 2013) and ice shelf disintegration (Massom et al., 2018). Due to these numerous interactions with the atmospheric boundary layer, cryosphere and upper ocean dynamics, ocean surface wave processes are becoming increasingly recognized as fundamental to climate on a range of spatial and temporal scales (e.g. Jenkins et al., 2012; Fan and Griffies, 2014; Qiao et al., 2016; Stoney et al., 2007; Aijaz et al., 2017; Cavaleri et al., 2012; Babanin et al., 2012; D'Asaro, 2012; Andreas et al., 2015; Stoney et al., 2018). In practical terms information about extreme waves is also critical for planning marine operations and the design of offshore marine infrastructure (Bell et al., 2017). By integrating the wind signal over large spatial scales (Barber and Ursell, 1948; Munk, 1963; Snodgrass, 1966), wind waves absorb and transmit the signature of synoptic-scale atmospheric dynamics (e.g. Gulev and Grigorieva, 2006; Semedo et al., 2015; Martínez-Asensio et al., 2016). In this respect, wind waves are an important indicator of climate variability and the intensity of atmospheric synoptic and mesoscale processes.

This is particularly true in the North Atlantic, which is characterized by high magnitude and strong variability in wind wave activity especially throughout the boreal winter (Gulev and Grigorieva, 2003, 2006; Semedo, 2011). This is the direct result of mid-latitude atmospheric baroclinicity in general and more specifically the vigorous flow that makes up the regional atmospheric storm track and the eddy-driven jet. Many authors have demonstrated a statistical association between interannual variability in wind wave climate and time-averaged atmospheric characteristics such as interannual fluctuations in the large-scale meridional pressure gradient referred to as the North Atlantic Oscillation (NAO; Bacon and Carter 1993; WASA Group, 1998; Gulev and Hasse, 1999; Wang and Swail, 2001; Woolf, 2002; Wang et al., 2004; Gulev and Grigorieva, 2006; Camus et al. 2014). There is a great deal of discussion regarding storm-track variability and so-called poleward deflection (e.g. Tamarin-Brodsky and Kaspi, 2017, Booth et al., 2017, Collins et al., 2018) which is found in some, but not all reanalyses (Tilinina et al., 2013) as well as some model simulations corresponding to warming climate scenarios (Pinto et al., 2007; Loeptien et al., 2008; Woollings and Blackburn, 2012). Uncertainty in changes of the local storm track and eddy driven jet propagate into projections of wave climate and limit our ability to have confidence in these diverse projections (Hemer et al., 2013; Fan et al., 2013; Khon et al., 2014; Markina and Gavrikov, 2016).

A meridionally asymmetric response is likely since the wave climate in the tropical, middle and high latitudes all fall under different influences and play roles in separate processes. In the tropics summertime wave heights are strongly impacted by changes in the intensity and frequency of tropical cyclones (e.g. Teague et al, 2007; Phibbs and Toumi, 2014) which are generally expected to become relatively sparser but more intense, and change in their zonal and meridional distribution over the coming century (e.g. Bender et al., 2010; Knutson et al., 2013; Studholme and Gulev, 2018). At higher latitudes, both winter and summer increases in seasonal-mean and maximum waves have been identified over the last 36 years (Waseda et al., 2018) and are projected to continue into the coming century (Khon et al., 2014; Casas-Prat et al., 2018). These regime changes are expected to result from a set of complex wave responses to both wind and sea-ice forcing acting in concert.

Wind wave climate variability is strongly associated with changes in wind speeds affected by cyclonic activity. In turn, atmospheric transient eddies may demonstrate different patterns of interannual variability from those of the mean winds and pressure gradients, especially locally (e.g. Gulev et al., 2002). The response of wind wave climate to atmospheric forcing is quite complex since waves in the open ocean are a composition of locally generated wind sea and remotely generated swell (Young et al, 2011). Thus, wind wave climate reflects local trends in wind speed as well as the frequency and intensity of atmospheric processes integrated over larger scales. Quantifying the responses of the wind wave climate to the differing impact of different spatial and temporal scales of atmospheric motions presented a considerable challenge. In this respect numerical wind wave modeling with advanced wave model configuration represents effective tools for simulations of wave characteristics as a function of a varying atmospheric forcing.

Here we analyze responses of surface wind waves to multi-scale atmospheric variability by looking at different bands of atmospheric wind variability and conducting a suite of numerical experiments to investigate the associated wind wave responses over boreal winter in the North Atlantic. This will help to derive further insights into the impact of storm track variability on wave climate and more specifically quantify the contribution of the atmospheric processes of different scales to the formation of mean and extreme wave characteristics over the boreal winter in the North Atlantic. In this way, uncertainty in wave climate projections may be constrained and better understood. The paper is organized as follows. Section 2 gives details of the model and its present configuration as well as data sets and analysis methodology used in this work. Section 3 presents the results of the analysis of the responses of modeled wind waves to the decomposed atmospheric forcing. Section 4 discusses the link between wave climate and atmospheric interannual variability at different scales. Section 5 summarizes the results and provides a brief discussion of the potential avenues for the further development of the study.

## 2. Numerical Simulations

*a. Wave model and experiment design*

Simulations are conducted with version 5.16 of the third-generation spectral wave model WAVEWATCH-III (WW3 herein; WW3DG, 2016) for the North Atlantic from 0 to 80° N and from 90°W to 15°E (Fig. 1). For this domain, the influence of swell originating south of the Equator is considered to be negligible and is thus ignored (e.g. Alves, 2006). In the WW3 setting, we use the source term package (ST4) parameterization for wave energy input and dissipation (Ardhuin et al., 2010) and the Discrete Interaction Approximation (DIA) scheme for non-linear wave interactions (Hasselmann and Hasselmann 1985). Model integration time step is 15 minutes. Simulations are performed for the boreal winter season (December-February; DJF) over the period 1980-2016. Individual model runs are initiated two weeks in advance (i.e. mid November) to account for the model spin-up. This initialization period is discarded from further calculations. The model settings described above have been used in a number of wave climate studies (e.g. Chawla et al., 2013, Rascle, Ardhuin, 2013).

Each seasonal experiment was run at 0.7° spatial resolution and forced by 10 m winds from the European Center for Medium-Range Weather Forecasts (ECMWF) ERA-Interim reanalysis (Dee et al., 2011). ERA-Interim is the demonstrably optimal reanalysis dataset to use for interannual studies of wave climate due to its relative consistency over the record period compared to alternative products (Stopa and Cheung, 2014). The ice source term package (IC0; Tolman, 2003) used in our WW3 configuration assumes the exponential attenuation of waves in partially sea-ice covered regions with further simple ice blocking based on 12-hourly ice concentrations from the ERA-Interim reanalysis.

Throughout this manuscript we use the notation, FULL, to refer to the results of experiment forced by original ERA-Interim winds. The FULL simulation serves as the reference model run. In three further experiments, wind forcing has been decomposed into three components corresponding to short-term sub-synoptic variability (0-2 days, hereafter referred to as SUBS), synoptic scale variability (here defined as 2-10 days; e.g. Hoskins and Hodges, 2002, hereafter referred to as SYNOP) and low frequency variability (more than 10 days, hereafter referred to as LF). Similar decomposition was earlier used by Ayrault et al. (1995) and Gulev et al. (2002). Wind field decomposition was performed using band-pass Lanczos filtering (Lanczos, 1956; Duchon, 1979) earlier effectively used by Gulev et al. (2002).

### b. *Diagnostics*

We concentrate on significant wave height ($H_s$) and mean wave direction ($\theta$), which are derived from the spectral model solution at grid points that are free from sea-ice for the whole period covered by the model integrations. See Liu et al. (2017) for the appropriate analytical expressions. In the analysis of co-variability between wave climate and atmospheric variability (section 4) we discuss mean and extreme characteristics ($95^{th}$ percentile). The maximum values of significant wave heights are discussed in section 3 act as a proxy for the upper bound of the obtained values and as is common for such estimations (e.g. Caires and Sterl, 2004; Janssen, 2015). This is defined as maximum values obtained from model simulations, averaged over the period 1980-2016 (DJF).

As an Eulerian measure of the intensity of atmospheric dynamic processes over a range of scales we use vertically integrated eddy kinetic energy (EKE) (Lorenz, 1955; Orlanski and Katzfey, 1991) computed from <2, 2-10 and >10 days bandpass-filtered 6-hourly wind fields (Blackmon, 1976; Blackmon et al., 1977; Hoskins and Hodges, 2002; Schneider, 2015; Woollings, 2016). The expression for EKE is given by

$$EKE = \int_{800}^{200} \left(\overline{u'^2} + \overline{v'^2}\right) dp/2g \qquad (1)$$

where $u'$ and $v'$ are bandpass-filtered zonal and meridional components of wind speed, $p$ is pressure (the vertical coordinate), and $g$ is acceleration due to gravity. We use band-pass filtering to isolate atmospheric transient eddies since there is a risk of inadvertently including stationary-eddies in more rudimentary eddy identification schemes (e.g. Yin, 2005; Mbengue et al., 2018). The remaining <2 days and >10 days band-passed flow act as high (sub-synoptic processes) and low frequency modes of atmospheric forcing. The integral is evaluated between 800 and 200 hPa to capture dynamical processes in the free troposphere (e.g. Mbengue et al., 2018). Since we are particularly interested in mid-latitude baroclinicity, a collection of potential metrics exist that could be used for this purpose. In particular in addition to vertically integrated EKE, near-surface eddy meridional heat fluxes ($v'\theta'$) and upper-tropospheric eddy momentum flux convergence ($-\nabla u'v'$) are both viable candidates as measures of the baroclinic-eddy life cycle (e.g. Chang et al., 2002; Mbengue and Schneider, 2013). Since here we are also interested in atmospheric dynamical processes more generally EKE is the logical candidate for the present analysis.

Fig. 2 shows the mean EKE for different band-passed ranges computed according to (1), and the cyclone track density in the winter season (defined here as DJF) 1980-2016 (Fig. 2b). The maximum values of EKE in synoptic range, $\sim 5.2 \cdot 10^5$ J m$^{-2}$, are located to the east of Newfoundland and are associated with the region of intensive cyclone genesis and development in the North Atlantic. The highest frequency mode (i.e. SUBS) of EKE (Fig.2a) demonstrates a spatial structure very similar to its synoptic-scale counterpart, though exhibits twice lower the magnitude. We note that by the nature of the bandpassing procedure, bundled in with true sub-synoptic dynamical processes, the SUBS filtering may also contain a small number of synoptic scale processes such as fast propagating cyclones (e.g. Rudeva and Gulev, 2011). The low frequency filtering of EKE (Fig. 2c) has the largest magnitudes (up to ~$30 \cdot 10^5$ J m$^{-2}$) with maximum values coinciding at the same location as for SUBS and SYNOP mode but in addition reveals an additional second maximum in the eastern subtropics. This EKE pattern (i.e. Eulerian measure) is highly consistent with the pattern of cyclone numbers (storm track density, i.e. a Lagrangian measure) in the North Atlantic also shown in Fig. 2b as derived from the ERA-Interim storm tracks provided by Kevin Hodges at University of Reading as used in Rogers (2014) based on methodology described in (Hoskins and Hodges, 2002, Hodges, 1995) and also exhibits high consistency with cyclone tracking climatologies available from the other numerical tracking algorithms applied to ERA-Interim (Neu et al. 2013; Rudeva and Gulev, 2011; Tilinina et al., 2013). There is tight correspondence between the Eulerian and Lagrangian characterizations of the storm-track in this particular region critical for wave formation (Gulev and Grigorieva, 2006). The Lagrangian tracking algorithm also appears to pick up a second local maxima off the south-eastern coast of Greenland which is presumably either larger polar low-type cyclones (Stoll et al, 2018) or cyclones associated with the Greenland tip jets (Vage et al., 2009) that are not seen in climatological EKE.

### 3. Wave climate responses to different scale atmospheric dynamical processes

*a. Climatologies*

The climatological seasonal-mean and seasonal-maximum distribution of significant wave heights and directions (DJF, 1980-2016) from the reference simulation (Fig. 1) shows maximum $H_s$ of ~4.7 m in the northeastern sector of North Atlantic consistent with Voluntary

Observing Ship (VOS) climatologies (Gulev and Hasse, 1998; Gulev et al. 2003), with satellite data (Zieger et al., 2009; Young et al., 2017) and with the ERA-Interim wave reanalysis (Dee et al., 2011). The pattern correlation with ERA-Interim wave reanalysis is ~0.97 for all winters (not shown), and the control experiment corresponds well to NDBC buoy data in the coastal areas (example for 2010 in shown in Fig. A1 in Appendix A). Mean $H_s$ is largest in an area displaced north-eastward from the maximum storm-track activity in the western North Atlantic (Fig. 1a). This is generally consistent with the mean direction and spatial scales of swell propagation, which both contribute largely to the total significant wave heights in this region (Chen et al., 2002; Semedo, 2011). In areas with the most intensive wind wave formation, such as the western tropics and the mid-latitudes, the dominant wave direction (red vectors) are more consistent with the mean wind direction (black vectors) than over the rest of the domain.

The spatial structure of maximum $H_s$ (Fig. 1b) reveals a pattern similar to the one observed for mean values (Fig. 1a) though it exhibits noisier structure, with several areas with maximal significant wave heights up to 15.9 m in the eastern mid-latitudes and local peak in the Labrador Sea with values up to 14 m. The observed spatial pattern to a degree reflects the influence of synoptic scale atmospheric structures on the formation of wave heights extremes.

Fig. 3 shows examples of the model snapshots for 1200 UTC 12/30/2000 as contained in the four abovementioned experiments (FULL, SUBS, SYNOP, LF) in which the simulated wave fields correspond directly to the features of atmospheric forcing of particular time scales. The smoothest and the most intense wind forcing field (LF) is reflected in large-scale wave patterns (Fig. 3c), while the synoptic (SYNOP) and sub-synoptic (SUBS) scale processes in the atmosphere imprint in the surface ocean on shorter spatial scales (Fig. 3a, b). The reference simulation with the full wind forcing (FULL) (Fig. 3d) superimposes the combined effect of the low and high frequency atmospheric forcing as would be expected.

The climatological seasonal-mean $H_s$ and $\theta$ distributions for all four simulations forced by the decomposed wind fields are presented in Fig. 4. The synoptic-scale processes (SYNOP), predominantly associated with cyclonic activity, are resolved into a spatial pattern with $H_s$ magnitudes being maximal (up to 2 m) in the mid-latitude North Atlantic (Fig. 4b). Sub-synoptic scale processes (SUBS) in general have the largest impact in the same areas as in SYNOP experiment, however, inducing waves which twice smaller (Fig. 4a). The low-frequency forcing (LF) dominates the wave distribution in the tropics (i.e. following from the existences of the steady flow of the easterly trade winds; Fig 4f) with $H_s$ values up to 2.5 m in the western tropical Atlantic (Fig. 4c). In addition, the LF component has a large impact over the eastern North Atlantic in mid-latitudes in the areas that have the largest wave heights in the reference model run (Fig. 1) with mean $H_s$ being up to 3 m. Therefore, the area of highest climatological $H_s$ (Fig. 1), while mostly dominated by LF, is also influenced by SUBS and SYNOP processes whose contribution is not negligible.

Both SUBS and SYNOP simulations demonstrate an almost purely divergent structure in $\theta$ emanating out from where the storm track intensity is maximal; this reflects their largely cyclonic origin (Fig. 4a, b and Fig. 2). The difference between these simulations and the control experiment (Figs. 4 d, e) reveals strong negative deviations everywhere with the largest differences along the eastern North Atlantic. The differences between LF and FULL experiments (Fig. 4f) are spread over the entire North Atlantic mid-latitudes, being lower compared to those in Figs. 4d,e. These spatial characteristics reveal the association between waves driven by synoptic and sub-synoptic scale processes and areas where the storm track is most active. They also emphasize the dominant role the low frequency atmospheric variability in forcing wind waves in the eastern North Atlantic mid-latitudes.

Fig. 5 shows the largest responses of wave heights on different atmospheric forcings in the same manner as Fig. 4 does for the mean values. Maximum wave heights presented in Fig. 5 a,b,. The patterns of maximum $H_s$ in SUBS, SYNOP and LF experiments are noisier compared to those for the mean values (Fig. 4). The highest extreme waves are identified in the Labrador Sea (in SYNOP) and in the Irminger Sea along the eastern coast of Greenland (in LF). These signatures are not present in the distribution of the mean $H_s$ (Fig. 4). The magnitudes of maximum wave heights for all simulations with decomposed forcing are comparable (8 m, 9.6 m, and 8 m for SUBS, SYNOP, and LF respectively). Physically this means that atmospheric motion across the entire range of temporal scales from sub-synoptic and synoptic transient eddies to lower frequency oscillations may provide an influence of equal magnitude upon ocean surface wave climate. The largest difference between the decomposed and reference simulations is observed in SUBS since this component initially has the lowest magnitudes of atmospheric forcing reflecting in the lowest values of simulated $H_s$ and amounts up to 16.5 m near the northern coast of the British Isles (Fig. 5d). For SYNOP and LF experiments the differences with FULL are up to 10 m (Fig. 5 e, f) and are observed over the northeastern North Atlantic.

The similar magnitudes of maximal $H_s$ in SUBS, SYNOP and LF simulations imply that the probability density distributions for $H_s$ have very different shapes across each experiment. In order to illustrate these regional connections between different scales of atmospheric variability on one hand and of wind wave heights on the other Fig. 6 shows histograms for $H_s$ at the sites indicated by yellow dots in Fig. 1. Stronger mid-latitude transient eddy activity (intensified storm track) leads to increasing waves along the North American Eastern seaboard, while it contributes less to the waves along the European coast since they are highly affected by low-frequency atmospheric forcing (LF). Note also, that transient eddies are contributing mostly to low magnitude waves on the eastern margin of the basin being up to 4 m near the coast of the British Isles (while the full range of $H_s$ expands up to 10 m), while on the western margin cyclone activity contributes to very large values to the whole spectrum of ocean surface wave distribution.

*b. Consideration of inherent nonlinearity: Aggregated decomposed-forcing climates compared to full-forced climate*

We note that the magnitudes of $H_s$ in the FULL simulations are close, but not exactly equal to the algebraic sum of the magnitudes of $H_s$ in the simulations forced by the decomposed wind flow separately (i.e. FULL ≈ SUBS + SYNOP + LF). This follows from the invocation of non-linear processes in wave growth and interactions. Moreover, Lanczos filtering used for wind forcing decomposition, while quite effective, may allow for the minor transfer of variance between the ranges (the so-called aliasing effect). The values for this misalignment (Fig. 7) vary from -90% to 58% with the average estimate being about -8.5 %. The mean total difference between the algebraic sum of simulations forced by decomposed wind fields and the magnitude of Hs in FULL is moderately negative while there are high-magnitude positive values found along the North American coast. Atmospheric and surface wave processes responsible for this pattern over the western margin of the basin vary depending on latitude. In the tropics this pattern likely results from the relatively smooth and consistent trade winds retained in LF. By contrast, in the mid-latitudes running up the eastern seaboard of the North American continent, this off-shore pattern is presumably dominated by transient eddies in the atmosphere in the SUBS and SYNOP, as the influence of LF is relatively weak in this region as it has been pointed out above (e.g. Fig.6 a and corresponding discussion).

Moderately negative values of the differences are widespread around the eastern margin of the basin and can presumably be explained by the fact that wave growth here results from the combination of a number of atmospheric processes operating in concert over a range of spatial and temporal scales (Fig. 6). Due to the exponential growth of wave energy at the initial stage of wave formation until the fully developed sea state when the wind wave spectrum is saturated (e.g. Miles, 1957; Cavaleri and Malanotte-Rizzoli, 1981) and thus an underestimation of [SUBS + SYNOP + LF] relative to FULL seems to be quite reasonable in the eastern mid-latitudes. In the eastern tropics and equatorial region, the difference is strongly negative and given generally low $H_s$ magnitudes here (less than 1 m in climatological seasonal mean, Fig.1a) it is presumably associated with the persistence of atmospheric forcing in LF simulations leading to formation of higher waves here relative to FULL experiment. The same is true for the Gulf of Mexico.

To summarize the role of different forcing components in forming mean and maximum $H_s$ we consider the ratio between wave heights in the experiments with decomposed forcing and the reference experiment (FULL) for the mean and the maximum $H_s$ (Fig. 7). This ratio will be further considered as a proxy for the upper bound of the observed contributions of different scales of the atmospheric dynamics in the wind wave field. Since the sum of the fractions is not equal 100% over the most of the area, this diagnostic complements the analysis of differences between FULL and decomposed forcing simulations presented above (Fig. 4 d,e,f, Fig. 5 d,e,f).

c. *Relative contribution to actual wave climate provided by different scales of atmospheric dynamical processes*

As mentioned above, the area with maximum seasonal-mean $H_s$ in the northeastern North

Atlantic is influenced by all three components of wind forcing, being up to 70%, 30 and 20% in LF, SYNOP and SUBS simulations respectively. However, if we consider the seasonal-maximum values of $H_s$ LS accounts for up to 60% with SYNOP and SUBS contributing 50 and 40% respectively. The area with the highest impact of synoptic-scale atmospheric variability (SYNOP, Fig. 8 b, e) is found to be located near the North American coast and in the Labrador Sea where waves forced by synoptic-scale winds can constitute up to 80% to mean and up to 90% to maximum significant wave heights. At the same time this area is to a lesser extent affected by low-frequency atmospheric variability (Fig. 8 c, f): the LF simulation show that waves have a general eastward direction (consistently with winds in LF) and thereby do not provide favorable conditions for fetch along the North American coast. While sub-synoptic scale atmospheric processes do not significantly contribute to the mean wave characteristics (Fig. 8 a), they do have a profound impact on the maximum waves (Fig. 8 d) over the main North Atlantic storm track area (Fig. 2) and in particular along the North American coast over the Gulf Stream.

In order to quantify the role of the different components to the total variability of wave heights, we analyze the ratio between the standard deviations ($\sigma$) in each experiment relative to the control simulation used as a proxy for their contributions. The largest values are observed in LF simulation (Fig. 9c) where standard deviation of $H_s$ has approximately the same magnitude as in FULL simulation, which reflects the dominant role LF forcing playing in waves formation in the tropics and in the semi-enclosed basins.

In the open ocean in the mid-latitudes, which is the main focus of this study, the majority of the total variability in $H_s$ is defined by combination of synoptic scale and low frequency forcing with differing impact in the eastern and western margin. Variability observed in LF simulations amounts up to 100% of the total variability in Hs along the British Isles while along the North American coast the major agent is synoptic scale forcing (up to 70% of the total variability; Fig. 9 b). Wave heights from the SUBS simulation demonstrate maximal variability compared to the control experiment in the semi-enclosed basins of the North Sea, Mediterranean, Gulf of Mexico and Gulf of Guinea and also has a local maximum along the North American coast. Given the fact that upper bound of the contribution to mean wave climate from the SUBS scales is very low relative to SYNOP and LF (Fig. 8a), and that its contribution is homogenous in the North Atlantic (Fig. 9a), we neglect this dynamical length-scale from the further analysis and concentrate on the response of wind wave climate on synoptic and low frequency modes of atmospheric forcing.

**4. Linking wave climate and atmospheric interannual variability at different scales.**

To study the large-scale atmospheric flow configurations invoking specific wind wave responses, we use synoptic (2-10 days) and low-frequency (>10 days) modes of EKE (Fig. 2) as a proxy for the intensity of the atmospheric dynamical processes with the largest impact on wave height formation in the North Atlantic. We explore their co-variability with $H_s$ and $\theta$ (EKE versus $H_s$ and EKE versus $\theta$) in the ocean basin. The dominant stationary modes of variability in the seasonal-mean (DJF) EKE are identified using EOF analysis applied to the de-trended time series (1979-2016) (Figure B1 in Appendix). Further we applied a canonical correlation analysis (CCA, von Storch and Zwiers, 1999) for mean values of vertically-integrated EKE (for both transient eddies and low frequency flow) and $H_s$ and for for two modes of EKE and $\theta$. The first five EOFs of EKE (2-10 days and >10 days) and $H_s$ were used for the CCA. The correlation coefficients between the first four modes of wave heights, mean wave direction and EKE (synoptic and low-frequency mode) are presented in Table 1. The correlation of the lead canonical pair is 0.90 for EKE synoptic mode (Fig. 10 a) and 0.95 for EKE low frequency mode (Fig. 10 c). The obtained spatial patterns for the synoptic mode of EKE are consistent with results of Lozano and Swail (2002). Since EKE canonical patterns with wave direction ($\theta$) have very similar spatial structure to the ones with wave heights, they are not presented here. Interestingly canonical patterns for both synoptic and low frequency modes of EKE reveals very similar locations of maximal absolute loadings of Hs which are further used to examine the associated wave heights regimes in the North Atlantic.

The first canonical pairs for both synoptic and low frequency modes of EKE implies that below-average ocean surface wave heights in the North-East Atlantic and above-normal ocean surface wave heights in the Eastern Atlantic are associated with a meridional displacement of the storm-track in the North Atlantic (Fig. 10 a, c). The major difference in the spatial structure of these canonical patterns comparing between transient eddies and low frequency variability is that the maximal loadings are strongly shifted across the basin eastward in EKE synoptic mode relative to the EKE low frequency mode. In general, the low frequency mode demonstrates a much more zonal spatial pattern than its synoptic counterpart. This is also true for the second canonical pattern discussed below. The first canonical patterns demonstrate that a northward shift of the storm-track corresponds to negative anomalies in wave heights in the Eastern North Atlantic (to the north-east from the Azores) and a southward shift of the storm-track is associated with positive wave anomalies in the North-East Atlantic to the east from Iceland, in the Norwegian and the North seas.

Unlike the similar first modes, the second canonical patterns of the synoptic and low-frequencies diverge from each other (Fig. 10 b, d). The pattern for the synoptic mode (fig. 10b) is presumably associated with storm-track intensity and indicates areas with positive loading in the storm-track region in the Eastern part of the North Atlantic associated with the negative loading in significant wave heights. In this way this pattern implies that a more intensive storm-track in the north-eastern part of the region results in wave height anomalies of the opposite sign. The region of negative EKE anomalies near the North American coast is characterized by nearly an order of magnitude weaker variability than the noted above region of positive anomalies of EKE and it is associated with wave height tendencies with the opposite sign. The observed pattern for the synoptic mode implies that areas with the strongest wave response to storm-track variability are shifted eastward and can be highly influenced by the processes taking place on the opposite site of the basin i.e. in the western North Atlantic mid-latitudes. Positive wave anomalies in the western and central North Atlantic (southward from Iceland) are closely associated with cyclone formation on the eastern margin of the basin. It stands to reason therefore that through this mechanism, storm track activity in the western North Atlantic is profoundly connected to wind wave anomalies along the North Atlantic eastern basin.

The second canonical pattern of EKE low frequency mode (Fig. 10 d) indicates the strengthening of the zonal flow reflecting in lower wave heights in the eastern North Atlantic mid-latitudes. In this way according to the second canonical patterns below-average wave heights in the North Atlantic eastern mid-latitudes can be considered either as a result of lower intensity of the storm-track or more intensive zonal flow in low frequency mode. CCA for the $95^{th}$ percentile of $H_s$ (not shown) reveals a very similar spatial pattern to the one observed for the mean values hence the above discussion is applicable to extreme waves as well.

The patterns for mean wave direction ($\theta$) demonstrate strong association with those observed for transient eddies and low-frequency flow (Fig. 10, shown in vectors). Above-average values of EKE are associated with eastward wave propagation, whereas below-average

EKE are associated with westward direction. Interestingly, the pattern for wave direction is much more coherent with modes of atmospheric variability while wave heights demonstrate the eastward displacement of maximum loadings relative to EKE.

Finally, to examine the specific response of wave heights in the areas with the largest association to synoptic and low frequency mode of the storm-track we compute the normalized occurrence anomalies of $H_s$ between 1980-2016 (Fig. 11) in the locations corresponding to the highest absolute values of canonical correlation patterns for significant wave heights from CCA with low-frequency forcing. In order to analyze modifying wave distribution depending on the regime associated with the certain mode of variability, we sort them as a function of each years rank in the first and second principal component time series from the $H_s$ EOFs. Values of the PC themselves are also shown (Fig. 11 c, f). Figures 11 a, b, d and e correspond to a 2-degree box with the center in sites indicated by purple and green dots in Fig. 10. These sites are objectively identified as the location of the minimum (Fig. 11 a, d) and maximum (Fig. 11 b, e) of the CCA pattern for $H_s$ for the first (Fig.11 a, b) and second (Fig. 10 d, e) PC respectively. During years with the lowest values of the first PC for $H_s$ (2010, 2001, 1996, 2013, 1982, 2011) an increase in the occurrence of high waves is observed in the Eastern part of the North Atlantic (area with maximum values of CCA patterns, Fig.11b) while at the same time a decrease in the occurrence of high waves is observed in the North-Eastern North Atlantic (minimum values of CCA patterns, Fig. 10a). The opposite pattern is observed during years with the highest values of the first PC (1993, 1989, 1981, 2015, 2012): in the North-Eastern part of the North Atlantic (near the Scandinavian coast, Fig. 11 a) waves become higher magnitude while in the Eastern North Atlantic (north-eastwards from Azores Islands, Fig. 11 b) the negative anomaly in the number of high waves is observed. For the second PC the picture is quite noisy for the Central North

Atlantic (Fig. 11 e), while for the North-Eastern North Atlantic (Fig. 11 d) the pattern is close to the one observed for the PC1 (Fig. 11 a): during years with highest values of PC2 (2014, 1990, 1994, 1995, 2016) higher than normal waves are observed, while moving toward the lowest values of PC2 we see the clear shift of waves distribution to lower values. Considering to extreme waves, increases/decreases in 95$^{th}$ percentile of $H_s$ are essentially consistent with the shift in wave distribution observed in occurrence anomalies (Fig. 11 a, b, d, e). However, as we see in the selected years (e.g. 1990 in Fig. 11 a, 1998 in Fig. 11 b, 2008 in Fig. 11 d), maximal wave heights in some way reflecting in the form of the distribution may not necessarily be captured in this shift of the wave regime since in general they are heavily dependent on the sampling variability.

## 5. Summary and discussion

We analyzed the differing responses of the ocean surface wind wave field in the North Atlantic to atmospheric dynamical processes of various scales in boreal winter. For this purpose, we have performed a suite of numerical experiments conducted with a state-of-the-art spectral wave model forced by band-pass filtered winds and thus dividing sub-synoptic (<2 days), synoptic (2-10 days), and low-frequency (>10 days) atmospheric forcing and specifically resolving the responses of wave climate to each of these individual components.

The region of seasonal-maximum wave height in the North Atlantic is displaced north-eastward of the area of the most vigorous tropospheric dynamics (measured here by vertically-integrated EKE with various bandpass-filtering). The sub-synoptic and synoptic-scale atmospheric forcings are found to have the largest impact upon wind waves along the North American coastline as well as in the Labrador Sea (up to 70% of total $H_s$). Meanwhile in the mid-latitudes (where mean wintertime $H_s$ is ~4.7 m), waves are generated by the superimposed contribution from both atmospheric synoptic-scale variability (likely associated with mobile cyclones) and lower frequency atmospheric forcing such as the prevailing westerlies. Sub-synoptic scale variability does not significantly contribute to seasonal-mean wind wave characteristics but does have considerable impact on the seasonal-maximum wave heights. This is particularly true along the North American coast and over the Gulf Stream.

In the subsequent analysis of responses to interannual atmospheric variability we concentrate on the influence of synoptic-scale transient eddies and low-frequency flow since they are found to have the largest effect on the absolute wave height and interannual variability. Interestingly, while wind waves in the eastern mid-latitudes of the North Atlantic are strongly influenced by low-frequency atmospheric forcing, with only a low estimated upper-bound on the contribution from synoptic scale variability (~30%), CCA of the dominant modes of synoptic scale variability in EKE and $H_s$ demonstrates that synoptic-scale processes in the western Atlantic are critically important for modulating wind wave variability in the eastern mid-latitudes. In this way the reduction in wave height along the European coast is likely associated with situations when weakening cyclone activity is observed over the North American Eastern Seaboard as well as with more intensive zonal flow within low-frequency mode. At the same time the meridional displacement of atmospheric transient eddies and low-frequency flow is associated with corresponding wave height anomalies. This relationship is also found to be reflected in the occurrence of anomalies of significant wave height distributions in sites along the eastern North Atlantic boundary. The shift in the wave regime, well captured by the leading principal components of $H_s$, reflects a shift in the ocean surface wave distribution and corresponding variability of extreme waves.

Here we have analyzed structural mechanisms driving wave climate responses to atmospheric variability rather than considering observed trends and signals. The conclusions presented in this study relate to significant wave height not accounting for the decomposition into wind sea and swell separately. Gulev and Grigorieva (2006) found that wind sea demonstrates the strongest association with local wind speed while swell is the most sensitive to the variation of cyclone counts. In this way analysis of the interannual variability of wave climate and the identified remote responses in the near-coastal areas requires consideration specifically in the context of potentially different signals in these characteristics.

Considering potential uncertainties in our results, it should be noted that all potential error and misrepresentation inherent in surface winds are transfused into wave models (Cavaleri, 2009) since momentum transfer from wind to ocean surface is the only energetic source for wave growth. Important issues exist with the various extrapolation schemes used in reanalysis products to yield a true surface layer from the lowest model level as well as with the various boundary layer formulations. Uncertainties in reanalysis surface winds are not limited to the boundary layer formulation in the atmospheric model, but may also include the impact of spatial and temporal inhomogeneities in data assimilation and inaccuracies in the computation of surface winds from the initial solution at model levels. Another source of uncertainty is associated with the interpolation of atmospheric forcing characteristics between the model time steps, particularly relevant in the highly turbulent atmospheric boundary layer. In these experiments atmospheric surface wind data are linearly interpolated from the initial 6-hourly time resolution of ERA-Interim output to the native integration time step of WW3 (15 minutes in this study). In this way higher temporal resolution of atmospheric forcing can provide a more detailed and reliable source of information for forcing any ocean wave model since, for example, the boundary layer particularities as well as atmospheric transient eddies can significantly alter in structure and location over the course of 6 hours (e.g. Held and Hoskins, 1985). Given this, the recently released ERA5 reanalysis with its 1-hourly temporal resolution and the recent version of NCEP-CFSv2 reanalysis (Saha et al., 2014) have promise for studying the significance of time interpolating surface winds. The use of higher resolution surface forcing functions, resolving mesoscale variability of winds, opens the route "inside cyclone" and allows for accounting for the role of mesoscale in forming wind wave variability. Nevertheless, since we are focused here on the effect of the different synoptic sub-ranges on the simulated wave climate, this source of uncertainty presumably does not have a significant impact on the results and we leave this sensitivity analysis for further work.

The analysis concept presented here provides an interesting avenue for the analysis of diversity in wind wave climate projections (e.g. Hemer et al., 2013, Wang et al., 2014, Aarnes et al., 2017). There is a great deal of discussion about various storm track and eddy driven jet changes under global warming. In the North Atlantic, climatological changes and interannual variability in the jet stream, and associated baroclinic transient eddies, can be described in some sense by a combination of the two leading large-scale patterns, the NAO and East Atlantic (EA) pattern (Woollings et al., 2010). These well-captured changes are associated with fluctuations in the jet stream's intensity and meridional displacement (Woollings et al., 2018). Taking into account the projected increase in the occurrence of positive phases of the NAO in the forthcoming century (e.g. Fan et al., 2013) as well as the evidence for an association between large-scale North Atlantic pressure gradients that resemble the NAO/EA with the geographical location of sea ice loss in the Arctic (Screen, 2017), storm track dynamics and associated transient eddies will likely undergo various changes relative to the contemporary climate in the near future (Ulbrich et al, 2013).

A poleward shift of the midlatitude storm track is one of the most widely discussed features in the observational records (Bender et al., 2012) and numerical model simulations of climate warming (Woollings et al., 2012; Bengtsson et al., 2006; Mbengue and Schneider, 2017) and can be mostly understood as a response to an alteration of the tropospheric meridional temperature gradient. Yin (2005) found that midlatitude westerlies also demonstrate a poleward shift. Ensemble mean model projections demonstrate evidence for increasing storm track activity in the eastern North Atlantic, amounting up to a 5-8% of increase in baroclinic wave activity by the end of the 21st century (Ulbrich et al., 2008, 2013). In addition, a poleward shift of the jet stream is observed (Woollings et al., 2012) as well as strengthening and poleward (and upward) shift of transient kinetic energy and momentum flux (Lorenz and DeWeaver, 2007), however considerable spread between models is found (Woollings et al., 2012, Zappa et al., 2013a).

In discussions of future climate projections, a considerable question remains of how well CMIP5 models represent the behavior of individual atmospheric transient eddies. For example, the majority of models demonstrate a reasonable number of extratropical cyclones, however in most of them the storm track is found to be either too zonally oriented or in some models to be displaced southward in the central North Atlantic (Zappa et al., 2013a). CMIP5 models also tend to underestimate cyclone intensity, specifically in the winter season (e.g. Zappa et al., 2013a). The magnitude of changes in storm-track intensity in the Northern Hemisphere is the largest in the eastern North Atlantic exceeding half of the interannual variability; this is found in up to 40% of CMIP5 models (Harvey et al., 2012), but again there is no consensus between models for the areas with maximum $H_s$ to the south from Iceland. This range of uncertainty, all of atmospheric origin, resoundingly influences future projections of wave climate, particularly in storm-track influenced areas in the North Atlantic (Hemer et al., 2013). In this way, understanding the differential effect of atmospheric flow decomposed into different lengthscales upon ocean surface waves can contribute a new perspective in understanding the future projections of the wind wave climate.

Ocean surface waves have recently been demonstrated to contribute significantly to interannual-to-multidecadal coastal sea-level changes (Melet et al., 2018). This work notes especially that the contribution of the wave climate changes to sea-level rise are both largely unconstrained and for the most part poorly appreciated. Given this, it is particularly relevant to observe regional responses of wave climate to variability in extratropical storm track activity. Ocean surface waves integrate signal from large-scale atmospheric phenomena and project this onto the regional scale. These results are therefore potentially useful for wave studies in both the open ocean and coastal areas.

*Acknowledgments*. We thank Cheikh Mbengue very much for a useful discussion, particularly regarding eulerian characterization of the storm track. We thank Kevin Hodges for providing data on cyclone trajectories and very much appreciate his comments on the first versions of manuscript. We thank Yulia Zyulyaeva for insights and a useful discussion of methods used for the analysis of obtained results. We also thank Alexander Babanin and Bernard Barnier for more general discussions on the first versions of the manuscript. This work was funded through the Agreement 14.W0331.006 with Ministry of Education and Science (analysis of forcing fields) and by the Russian Federation and Grant 14-50-00095 from the Russian Science Foundation (model simulations).

**APPENDIX A. Validation of reference simulations**

In the main body of the paper we claim a good agreement between the results of model experiments and significant wave heights from ERA-Interim reanalysis (Section 2a) and mentioned that model configuration has been widely used in a variety of wave climate studies (e.g. Chawla et al., 2013, Rascle, Ardhuin, 2013). However, as discussed above, coastal areas are particularly difficult to model with high veracity and they are known to under-estimate higher magnitude wave heights there in particular (Stopa and Cheung, 2014). Figure A1 shows the validation against high-resolution buoy data available from NDBC (http://www.ndbc.noaa.gov/) along the US eastern seaboard and demonstrate mean Pierson correlation coefficient being 0.92 (varying from 0.9 to 0.95). In view of the arguments presented above model results are considered to be reliable for the analysis conducted in this study.

**APPENDIX B. Eddy Kinetic Energy and Significant Wave Height EOFS**

Figure B1 shows the EOFS for seasonal-mean EKE SYNOP and LF, $H_s$ and $\theta$. The presented patterns are similar to the ones obtained for canonical patterns between these parameters. EOF analysis shows that both the synoptic and low frequency modes of EKE reveal similar patterns with $H_s$ with significant portion of the total variance being contained within the first two EOFs: the first mode contains 40.6% and 31.5% of variability in LF and SYNOP EKE respectively and 43% of the variability in significant wave height. The second EOF corresponds to 26.2% and 17.9% of variability in LF and SYNOP EKE respectively and 30.7% of variability in $H_s$ with smaller values for the subsequent modes. Regarding wave direction, the first EOF accounts for 29.5% and the second EOF for 14.5% of variability in $\theta$. The correlation coefficient between the first principal components of $H_s$ and EKE is 0.78 for EKE synoptic mode and 0.92 for EKE low frequency mode ($p$-value < 0.00001 in both cases), which confirms the link between these two characteristics.

**References**

Aarnes, O. J., M. Reistad, Ø. Breivik, E. Bitner-Gregersen, L. Ingolf Eide, O. Gramstad, A. K. Magnusson, B. Natvig, and E. Vanem (2017), Projected changes in significant wave height toward the end of the 21st century: Northeast Atlantic, J. Geophys. Res. Oceans, 122, 3394–3403, doi:10.1002/2016JC012521.

Aijaz, S., M. Ghantous, A. V. Babanin, I. Ginis, B. Thomas and G. Wake, 2017: Non-breaking wave-induced mixing in upper ocean during tropical cyclones using coupled hurricane-ocean-wave modeling. *Journal of Geophysical Research - Oceans*. **122**. 3939-3963.

Alves, J.-H. G.M., 2006: Numerical modeling of ocean swell contributions to the global wind-wave climate, *Ocean Modelling*, **11**, pp. 98–122.

Andreas, E. L., L. Mahrt, and D. Vickers, 2015: An improved bulk air-sea flux algorithm, including spray-mediated transfer. *Quarterly Journal of the Royal Meteorological Society*. **141**. 642-654.

d'Asaro, E., 2012: Turbulence in the upper-ocean mixed layer. *Annu. Rev. Mar. Sci.* **6**, 101-115.

Ardhuin, F., W. E. Rogers, A. V. Babanin, J. Filipot, R. Magne, A. Roland, A. van der Westhuysen, P. Queffeulou, J. Lefevre, L. Aouf and F. Collard, 2010: Semiempirical dissipation source functions for ocean waves. Part I: Definition, calibration, and validation. *J. Phys. Oceanogr.*, **40**, 1,917–1,941.

Babanin, A. V., 2006: On a wave-induced turbulence and a wave-mixed upper ocean layer. *Geophys. Res. Lett.*, **33**, L20605, doi:10.1029/2006GL027308.

Babanin, A. V., A. Ganopolski, and W. R. C. Phillips, 2009: Wave-induced upper-ocean mixing

in a climate model of intermediate complexity. *Ocean Modell.*, **29**, 189–197.

Babanin, A. V., 2011: *Breaking and Dissipation of Ocean Surface Waves*. Cambridge University Press. 480 pp.

Babanin, A., M. Onorato, and F. Qiao, 2012: Surface waves and wave-coupled effects in lower atmosphere and upper ocean. *Journal of Geophysical Research.* **117.**

Bacon, S., Carter, D.J.T., 1993: A connection between mean wave height and atmospheric pressure gradient in the North Atlantic, *International Journal of Climatology*, **13**, 4, pp.423-436.

Barber, N.F., Ursell F., 1948: The generation and propagation of ocean waves and swell.I. Wave periods and velocities. *Philosophical transactions of the Royal Society A. Mathematical, Physical and Engineering Sciences*.

Bell, R.J., Grey, S.L., Jones, O.P., 2017: North Atlantic storm driving of extreme wave heights in the North Sea, Journal of Geophysical Research: Oceans, **122**, 4

Bengtsson, L., K. Hodges, and E. Roeckner, 2006: Storm tracks and climate change. *J. Climate*, **19**, 3518–3543, doi:10.1175/ JCLI3815.1.

Bender, M. A., T. R. Knutson, R. E. Tuleya, J. J. Sirutis, G. A. Vecchi, S. T. Garner, and I. M. Held, 2010: Modeled Impact of Anthropogenic Warming on the Frequency of Intense Atlantic Hurricanes. *Science*. **327**. 454-458.

Bender, F., V. Ramanathan, and G. Tselioudis, 2012: Changes in extratropical storm track cloudiness 1983-2008: observational support for a poleward shift. *Climate Dynamics*, **38**, 2037–2053.

Blackmon M (1976) A climatological spectral study of the 500 mb geopotential height of the Northern Hemisphere. *J. Atmos. Sci.* **33**. 1607–1623.

Blackmon M, Wallace J, Lau N, Mullen S (1977) An observational study of the Northern Hemisphere wintertime circulation. *J. Atmos. Sci.* **34.** 1040–1053.

Booth, J.F. 2017: Spatial Patterns and Intensity of the Surface Storm Tracks in CMIP5 Models, Journal of Climate, 30, 4965-4981.

Caires, S., Sterl, A. 2004: 100-Year Return Value Estimates for Ocean Wind Speed and Significant Wave Height from the ERA-40 Data, Journal of Climate, 18, 1032-1048.

Camus, P., Menéndez, M., Méndez, F. J., Izaguirre, C., Espejo, A., Cánovas, V., Medina, R. (2014). A weather-type statistical downscaling framework for ocean wave climate. *Journal of Geophysical Research: Oceans,* **119** (11), 7389-7405. 10.1002/2014JC010141.

Casas-Prat, M., X. L., Wang, and N. Swart, 2018: CMIP5-based global wave climate projections including the entire Arctic Ocean. *Ocean Modelling.* **123**, 66-85.

Cavaleri, L. and P. Malanotte-Rizzoli, 1981: Wind-wave prediction in shallow water: Theory and applications. *J. Geophys. Res.*, **86**, 10,961–10, 973.

Cavaleri, L., 2009: Wave Modeling—Missing the Peaks, Journal of Physical Oceanography, v.39, pp.2757-2778.

Cavaleri, L. B. Fox-Kemper, and M. Hemer, 2012: Wind waves in the coupled climate system. *Bulletin of the American Meteorological Society.* **93**. 1651-1661.

Chang, E. K. M., S. Lee, and K. L. Swanson, 2002: Storm Track Dynamics, *J. Clim*. **15**. 2163 - 2183.

Chen, G., Chapron, B., Ezraty, R., 2002: A Global View of Swell and Wind Sea Climate in the Ocean by Satellite Altimeter and Scatterometer, *J. Atmos. Ocean. Tech.*, **19**, 1849-1859.

Collins, M., et al. (2018). Challenges and opportunities for improved understanding of regional climate dynamics. Nature Climate Change, 8(2), 101-108. doi:10.1038/s41558-017-0059-8

Dee, D., P., S. M. Uppala, A. J. Simmons, P. Berrisford, P. Poli, S. Kobayashi, U. Andrae, M. A. Balmaseda, G. Balsamo, P. Bauer, P. Bechtold, A. C. M. Beljaars, L. van de Berg, J. Bidlot, N. Bormann, C. Delsol, R. Dragani, M. Fuentes, A. J. Geer, L. Haimberger, S. B. Healy, H. Hersbach, E. V Hólm, L. Isaksen, P. Kållberg, M. Köhler, M. Matricardi, A. P. McNally, B. M. Monge-Sanz, J.-J. Morcrette, B.-K. Park, C. Peubey, P. de Rosnay, C. Tavolato, J.-N. Thépaut, F. Vitart. 2011: The ERA-Interim reanalysis: configuration and performance of the data assimilation system, *Q. J. R. Meteorol. Soc.* **137**(656). P. 553–597.

Duchon, C.E. 1979. Lanczos filtering in one and two dimensions. J. Appl. Meteorol. 18: pp. 1016 – 1022.

Fan, Y., Held, I.M., Lin S.-J., Wang, X. L., 2013: Ocean Warming Effect on Surface Gravity Wave Climate Change for the End of the Twenty-First Century. *Journal of Climate*, **26**, 6046-6066.

Fan, Y. and S. Griffies, 2014: Impacts of parameterized langmuir turbulence and nonbreaking wave mixing in global climate simulations. *J. Clim*. **27**. 4752-4775.

Gulev, S.K., T. Jung, and E. Ruprecht, 2002: Climatology and interannual variability in the intensity of synoptic-scale processes in the North Atlantic from the NCEP-NCAR Reanalysis data. *J.Clim.*, 15, 809-828.

Gulev, S.K., Grigorieva, V., Sterl, A., Woolf, D., 2003: Assessment for the reliability of wave observations from Voluntary Observing Ships: insights from the validation of a global wind wave climatology based on voluntary observing ship data. *J. Geophys. Res. – Oceans,* **108**(C7), p. 3236, doi:10,1029/2002JC001437.

Gulev, S.K., Grigorieva, V., 2006: Variability of Winter Wind Waves and Swell in the North Atlantic and North Pacific as Revealed by the Voluntary Observing Ship Data, *Journal of Climate,* **19**, pp. 5667-5685.

Gulev, S. K., and L. Hasse, 1999: Changes of wind waves in the North Atlantic over the last 30 years, *Int. J. Climatol.*, **19**, 1018–1091.

Harvey, B.J., Shaffrey, L.C., Woollings, T.J., Zappa, G., Hodges, K.I., 2012: How large are projected 21st century storm track changes? Geophysical Research Letters, **39**, L18707, doi:10.1029/2012GL052873

Hasselmann, S., and K. Hasselmann. 1985: Computations and parameterizations of the nonlinear energy transfer in a gravity-wave spectrum. Part I: A new method for efficient computations of the exact nonlinear transfer integral, *J. Phys. Oceanogr.*, **15**, 1369–1377.

Held, I. M., and B. J., Hoskins, 1985: Large-scale eddies and the general circulation of the troposphere. *Adv. Geophysics.* **28A**, 3-31.

Hemer, M. A, Fan, Y., Mori, N., Semedo, A., Wang, X.L., 2013: Projected changes in wave climate from a multi-model ensemble, *Nat. Clim. Change*, **3**, 471–476.

Hodges K.I. 1995: Feature tracking on the unit sphere. *Mon. Wea. Rev.* **123**(12): 3458–3465.

Hoskins B., Hodges K. (2002) New perspectives on the Northern Hemisphere winter storm

tracks. *J. Atmos. Sci.* **59**. 1041–1061.

Janssen, P. A. E. M., 2015: Notes on the maximum wave height distribution, ECMWF Technical Memorandum No. 755, 21 p.

Jenkins, A., M. Paskyabi, I. Fer, A. Gipta, and M. Adakudlu, 2012: Modelling the effect of ocean waves on the atmospheric and ocean boundary layers. *Energy Procedia.* **24**. 166-175.

Khon, V. C., I. I. Mokhov, F. A. Pogarsky, A. Babanin, K. Dethloff, A. Rinke, and H. Matthes. 2014: Wave heights in the 21st century arctic ocean simulated with a regional climate model. *Geophysical Research Letters.* **41**, 2956-2961.

Knutson, T. R., J. J. Sirutis, G. A. Vecchi, S. Garner, M. Zhao, H.-S. Kim, M. Bender, R. E. Tuleya, I. M. Held, and G. Villarini, 2013: Dynamical downscaling projections of 21st century Atlantic hurricane activity: CMIP3 and CMIP5 model-based scenarios. *Journal of Climate.* **26**, 6591-6617.

Lanczos, C., 1956: Applied Analysis. Prentice-Hall, 539 pp

de Leeuw, G., E. L Andreas, M. D. Anguelova, C. W. Fairall, E. R. Lewis, C. O'Dowd, M. Schulz, and S. S. Schwartz, 2011: Production flux of sea spray aerosol. *Rev. Geophys.*, **49**, RG2001, doi:10.1029/2010RG000349.

Liu, Q., A. Babanin, Y. Fan, S. Zieger, C. Guan, and I.-J. Moon, 2017: Numerical simulations of ocean surface waves under hurricane conditions: Assessment of existing model performance. *Ocean Modelling.* **118**. 73-93.

Lorenz, D.J. DeWeaver, E.T., 2007: Tropopause height and zonal wind response to global warming in the IPCC scenario integrations. *J. Geophys. Res.*, 112, D10119.

Lozano, I., Swail V., 2002: The link between wave height variability in the north Atlantic and the storm track activity in the last four decades, *Atmosphere-Ocean*, **40**(4) 377.

Markina, M. Yu., and A. V. Gavrikov, 2016: Wave Climate Variability in the North Atlantic in Recent Decades in the Winter Period Using Numerical Modeling. *Oceanology.* **56**, 320-325.

Martínez-Asensio, A., Tsimplis, M. N., Marcos, M., Feng, X., Gomis, D., Jordà, G., & Josey, S. A. (2016). Response of the north atlantic wave climate to atmospheric modes of variability. *International Journal of Climatology,* **36**(3), 1210-1225. 10.1002/joc.441.

Massom, R. A., T. A. Scambos, L. G. Bennets, P. Reid, V. A. Squire, and S. E. Stammerjohn, 2018: Antarctic ice shelf disintegration triggered by sea ice loss and ocean swell. *Nature.* **558**. 383-389.

Mbengue, C., and T. Schneider, 2013: Storm Track Shifts under Climate Change. What Can Be Learned from Large-Scale Dry Dynamics. *J. Clim.* **26**. 9923 - 9930.

Mbengue, C., Schneider, T., 2017: Storm-Track Shifts under Climate Change: Toward a Mechanistic Understanding Using Baroclinic Mean Available Potential Energy, *J. Atmos. Sci*, **74**, pp. 93-110.

Mbengue, C.O., Woollings, T., Dacre, H.F. et al. 2018: The roles of static stability and tropical–extratropical interactions in the summer interannual variability of the North Atlantic sector, *Clim. Dyn*, pp. 1-17.

Melet, A., Meyssignac, B., Almar, R., Le Cozannet, G., 2018: Under-estimated wave contribution to coastal sea-level rise. *Nature Climate Change*, **8**, 234-239.

Miles, J.M., 1957: On the generation of surface waves by shear flows, *Journal of Fluid Mechanics*, 3, 2, 185-204.

Munk, W. H., G. R. Miller, F. E. Snodgrass, and N. F. Barber, 1963: Directional recording of swell from distant storms. *Philos. Trans. Roy. Soc. London*, A255, 505–584.

Orlanski, I. and J. Katzfey, 1991: The Life Cycle of a Cyclone Wave in the Southern Hemisphere. Part I: Eddy Energy Budget. *J. Atmos. Sci.* **48**:17. 1973-1998.

Phibbs, S. and R. Toumi, 2014: Modeled Dependance of Wind and Waves on Ocean Temperature in Tropical Cyclones. *Geophysical Research Letters*. **41**, 7383-7390.

Pinto, J. G., U. Ulbrich, G. C. Leckebusch, T. Spangehl, M. Reyers, and S. Zacharias, 2007: Changes in storm track and cyclone activity in three SRES ensemble experiments with the ECHAM5/MPI-OM1 GCM. *Climate Dynamics*. **29**, 195–210.

Qiao, F., Y. Yuan, J. Deng, D. Dai, and Z. Song, 2016: Wave-turbulence interaction-induced vertical mixing and its effects in ocean and climate models. *Philosophical transactions of the Royal Society A.*

Rascle N, Ardhuin F (2013) A global wave parameter database for geophysical applications. Part 2: Model validation with improved source term parameterization. *Ocean Modelling.* doi: 10.1016/j.ocemod.2012.12.001

Roberts, J. F., Champion, A. J., Dawkins, L. C., Hodges, K. I., Shaffrey, L. C., Stephenson, D. B., Stringer, M. A., Thornton, H. E., and Youngman, B. D.: The XWS open access catalogue of extreme European windstorms from 1979 to 2012, *Nat. Hazards Earth Syst. Sci.*, 14, 2487-2501, https://doi.org/10.5194/nhess-14-2487-2014, 2014.

Rudeva, I., and S.K. Gulev, 2011: Composite analysis of the North Atlantic extratropical cyclones in NCEP/NCAR reanalysis. *Mon. Wea. Rev.*, **139**, pp. 1419-1436.

Saha, S., ;S. Moorthi, X. Wu, J. Wang, and Coauthors, 2014: The NCEP Climate Forecast System Version 2. *J. Clim.*, 27, 2185–2208, doi:10.1175/JCLI-D-12-00823.1

Schneider T., Bischoff T., Potka H. (2015) Physics of changes in synoptic midlatitude temperature variability. *J Clim.,* **28**(6):2312–2331.

Screen, J.A., 2017: Simulated atmospheric response to regional and pan-Arctic sea ice loss. *J Clim.,*, **30**, 3945-3962.

Semedo, A., Vettor, R., Breivik, Ø., Sterl, A., Reistad, M., Soares, C. G., & Lima, D. (2015). The wind sea and swell waves climate in the nordic seas. *Ocean Dynamics,* **65**(2), 223-240. 10.1007/s10236-014-0788-4.

Semedo, A., Sušelj, K., Rutgersson, A., Sterl, A., 2011: A Global View on the Wind Sea and Swell Climate and Variability from ERA-40, *J Clim.,* **24**(5), 1461-1479.

Snodgrass, F. E., G. W. Groves, K. F. Hasselmann, G. R. Miller, W. H. Munk, and W. M. Powers, 1966: Propagation of swell across the Pacific. *Philos. Trans. Roy. Soc. London*, A259, 431–497.

Stoll, P., Graversen, R. G., Noer, G. and Hodges, K. 2018: An objective global climatology of polar lows based on reanalysis data. *Quarterly Journal of the Royal Meteorological Society*. ISSN 1477-870X (In Press).

Stopa, J.E., and K. F. Cheung, 2014. Intercomparison of Wind and Wave Data from the ECMWF Reanalysis Interim and the NCEP Climate Forecast System Reanalysis. *Ocean Modelling*,

75, 65-83. doi:10.1016/j.ocemod.2013.12.006

Stoney, L., K. Walsh, A. V. Babanin, M. Ghantous, P. Govekar, and I. Young, 2007: Simulated ocean response to tropical cyclones: The effect of a novel parameterization of mixing from unbroken surface waves. *Journal of Advances in Modeling Earth Systems.* **9**, 759-780.

Stoney, L., K. J. E. Walsh, S. Thomas, P. Spence and A. V. Babanin, 2018; Changes in Ocean Heat Content Caused by Wave-Induced Mixing in a High-Resolution Ocean Model. *Journal of Physical Oceanography.* **48**, 1139-1150.

Studholme, J. H. P., and S. K. Gulev, 2018: Concurrent Changes to Hadley Circulation and the Meridional Distribution of Tropical Cyclones. *J Clim.,* **31**, 4367-4389.

Sullivan, P. P., and J. C. McWilliams, 2002: Turbulent flow over water waves in the presence of stratification. *Phys. Fluids*, **14**, 1182–1195.

Tamarin-Brodsky, T., & Kaspi, Y. (2017). Enhanced poleward propagation of storms under climate change. Nature Geoscience, 10(12), 908-913. doi:10.1038/s41561-017-0001-8

Teague, W. J., E. Jarosz, D. W., Wang, and D. A. Mitchell, 2007: Observed Oceanic Response over the Upper Continental slope and Outer Shelf During Hurricane Ivan. *Journal of Physical Oceanography.* **37**, 2181-2206.

Tilinina, N., Gulev, S. K., Rudeva, I., & Koltermann, P. (2013). Comparing cyclone life cycle characteristics and their interannual variability in different reanalyses. *J Clim., 26*(17), 6419-6438.

Tolman, H.L. (2003). Treatment of unresolved islands and ice in wind wave models, *Ocean Modell.*, 5, pp. 219-231

Ulbrich, U., Pinto, J. G., Kupfer, H., Leckebusch, G. C., Spangehl, T. and Reyers, M. 2008 Changing northern hemisphere storm tracks in an ensemble of IPCC climate change simulations *J. Clim.* **21** 1669–79.

Ulbrich, U., et al, 2013: Are Greenhouse Gas Signals of Northern Hemisphere winter extra-tropical cyclone activity dependent on the identification and tracking algorithm? *Meteorologische Zeitschrift*, **22** (1), p. 61 – 68

Vage, K., Spengler, T., Davies H.C., Pickart, R.S., 2009: Multi-event analysis of the westerly Greenland tip jet based upon 45 winters in ERA-40, Q. J. R. Meteorol. Soc. 135: 1999–2011.

Veron, F., W. K. Melville, and L. Lenain, 2008: Wave-coherent air–sea heat flux. *J. Phys. Oceanogr.*, **38**, 788–802.

Veron, F., W. K. Melville, and L. Lenain, 2011: The effects of small-scale turbulence on air–sea heat flux. *J. Phys. Oceanogr.*, **41**, 205–220.

von Storch, H., and F. W. Zwiers, 1999: Statistical Analysis in Climate Research. Cambridge University Press, 503 pp

Wang, X.L., Zwiers, F.W., Swail, V.R., 2004: North Atlantic Ocean Wave Climate Change Scenarios for the Twenty-First Century, *J Clim.,* **17**, pp. 2368-2383.

Wang, X. L., and V. R. Swail, 2001: Changes of extreme wave heights in Northern Hemisphere oceans and related atmospheric circulation regimes, *J. Clim.*, **14**, 2201–2204.

Wang, X. L., Y. Feng, and V. R. Swail (2014), Changes in global ocean wave heights as projected using multimodel CMIP5 simulations, Geophys. Res. Lett., 41, 1026–1034,

doi:10.1002/2013GL058650.

WASA Group, 1998: Changing waves and storms in the northeast Atlantic, *Bull. Am. Meterol. Soc.,* **79**, 741–760.

WAVEWATCH III Development Group (WW3DG), 2016: User manual and system documentation of WAVEWATCH III R version 5.16. Tech. Note 329, NOAA/NWS/NCEP/MMAB, College Park, MD, USA, 326 pp. + Appendices.

Woolf, D., P. Challenor, and P. Cotton, 2002: Variability and predictability of the North Atlantic wave climate. *J. Geophys. Res.*, **107**, 3145, doi:10.1029/2001JC001124.

Woollings T , Hannachi A, Hoskins B. 2010. Variability of the North Atlantic eddy-driven jet stream. Q. J. R. Meteorol. Soc. 136: 856–868. DOI:10.1002/qj.625

Woollings, T., and M. Blackburn, 2012: The North Atlantic jet stream under climate change and its relation to the NAO and EA patterns. *J Clim.*, 25, 886–902.

Woollings, T., Gregory, J.M., Pinto, J.G., Reyers, M., Brayshaw, D.J., 2012: Response of the North Atlantic storm track to climate change shaped by ocean-atmosphere coupling, *Nature Geoscience*, **5**, 313-317.

Woollings T, Papritz L, Mbengue C, Spengler T (2016) Diabatic heating and jet stream shifts: a case study of the 2010 negative North Atlantic oscillation winter. *Geophys. Res. Lett.* **43**(18): 9994–10002. https://doi.org/10.1002/2016GL070146.

Woollings, T., E. Barnes, B. Hoskins, Y-O. Kwon, R. W. Lee, C. Li, E. Madonna, M. McGraw, T. Parker, R. Rodrigues, C. Spensberger, and K. Williams, 2018: Daily to Decadal Modulation of Jet Variability. *J. Clim.* **31**. 1297-1314.

Yin, J. H., 2005: A consistent poleward shift of the storm track in simulations of 21st century climate. *Geophys. Res. Lett.*, **32**, L18701, doi:10.1029/2005GL023684.

Young, I.R., Zieger, S., Babanin, A.V., 2011: Global trends in wind speed and wave height. *Science*. 332(6028), 451-5.

Young I, Sanina E, Babanin A., 2017: Calibration and Cross Validation of a Global Wind and Wave Database of Altimeter, Radiometer, and Scatterometer Measurements. *Journal of Atmospheric and Oceanic Technology*, **34,** 6. DOI: 10.1175/JTECH-D-16-0145.1.

Zappa, G., Shaffrey, L.C., Hodges K.I., 2013a: The Ability of CMIP5 Models to Simulate North Atlantic Extratropical Cyclones, *J Clim.,* **26**, 5379-5396.

Zappa, G., Shaffrey, L. C., Hodges, K. I., Sansom, P. G. and Stephenson, D. B., 2013b: A Multimodel Assessment of Future Projections of North Atlantic and European Extratropical Cyclones in the CMIP5 Climate Models, *J Clim.,* **26**, 5846-5862.

Zieger, S., Vinoth, J. and Young, I. R., 2009: Joint Calibration of Multi-Platform Altimeter Measurements of Wind Speed and Wave Height Over the Past 20 Years. *J. Atmos. Ocean. Tech.*, **26**(12), 2549-2564.

**LIST OF TABLES**

**Table 1.** Canonical correlation coefficients for the first four canonical pairs of wave heights and wave direction and synoptic and low-frequency modes of EKE.

Table 1. Canonical correlation coefficients for the first four canonical pairs of wave heights and wave direction and synoptic and low-frequency modes of EKE.

| Canonical Correlation | 1 | 2 | 3 | 4 |
|---|---|---|---|---|
| $H_s$ and EKE(LF) | 0.95 | 0.86 | 0.83 | 0.65 |
| $H_s$ and EKE(SYNOP) | 0.90 | 0.75 | 0.52 | 0.54 |
| $\theta$ and EKE(LF) | 0.96 | 0.95 | 0.85 | 0.76 |
| $\theta$ and EKE(SYNOP) | 0.85 | 0.77 | 0.65 | 0.65 |

# LIST OF FIGURES

**Fig. 1.** Simulated climatological seasonal-mean (a) and seasonal-maximum (b) significant wave heights (shading), mean wave and wind directions (red and black vectors respectively) for DJF 1980-2016 (FULL). Colored dots correspond to buoy sites used in further analysis.

**Fig. 2.** Climatological seasonal-mean vertically integrated eddy kinetic energy [$\cdot 10^5$ J m$^{-2}$] bandpass-filtered for <2days (a), 2-10 days (b) and >10 days (c) and cyclone track density per season (DJF) per 2 degree box in 1980-2016 (shown in shading on (b)).

**Fig. 3.** Snapshots of significant wave heights simulated by the decomposed wind fields (a) SUBS, (b) SYNOP, (c) LF and full forcing (d) FULL at UTC 12:00 2000-12-30.

**Fig. 4**. Climatological seasonal mean significant wave heights forced by decomposed wind fields: $H_s$ from SUBS (a), SYNOP (b) and LF (c) simulations and their difference with full forcing simulations: (d) [SUBS-FULL]; (e) [SYNOP-FULL]; (f) [LF-FULL].

**Fig. 5.** Maximum significant wave heights forced by decomposed wind fields:
$H_s$ from SUBS (a), SYNOP (b) and LF (c) simulations and their difference with full forcing simulations: (d) [SUBS-FULL]; (e) [SYNOP-FULL]; (f) [LF-FULL].

**Fig. 6.** Histograms of wave heights from SUBS, SYNOP, LF and FULL wave model simulations on the eastern (a) and western (b) margins of the North Atlantic sited in red dots in Fig.. 1a.

**Fig.. 7.** Difference between sum of wave heights simulated by decomposed forcing and waves simulated by full forcing (i.e. $[\text{SUBS} + \text{SYNOP} + \text{LF}] - [\text{FULL}]$).

**Fig.. 8.** The ratio between mean (a,b,c) and maximum (d,e,f) significant wave heights from simulations with decomposed forcing and reference simulations:
$\frac{[\text{SUBS}]}{[\text{FULL}]} * 100\%$ (a),(d); $\frac{[\text{SYNOP}]}{[\text{FULL}]} * 100\%$ (b),(e); $\frac{[\text{LF}]}{[\text{FULL}]} * 100\%$ (c),(f).

**Fig. 9.** Ratio between standard deviations of significant wave heights from simulations with decomposed forcing and reference simulations: $\frac{\sigma[\text{SUBS}]}{\sigma[\text{FULL}]} * 100\%$ (a); $\frac{\sigma[\text{SUBS}]}{\sigma[\text{FULL}]} * 100\%$ (b); $\frac{\sigma[\text{LF}]}{\sigma[\text{FULL}]} * 100\%$ (c).

**Fig. 10.** Pairs for the first two CCA modes between mean $H_s$ and modes of atmospheric variability: 2-10 days band-passed filtered i.e. SYNOP (a,b), and >10 days band-passed filtered i.e. LF, vertically integrated EKE (c,d); a,c represent the first and b,d the second CCA respectively. Solid (dashed) lines indicate positive (negative) values of EKE. Red (blue) scale represents positive (negative) values of the $H_s$ field. Dots are positioned at the locations of maximum (blue) and minimum (purple) values of CCA pattern for $H_s$ and EKE low frequency mode. Vectors indicate CCA modes for wave heights direction (θ).

**Fig. 11.** Normalized occurrence anomalies of $H_s$ in DJF 1980-2016 in the 2-degrees boxes with the centers shown in purple(a)/blue(b) dots in Fig.. 10 a and purple(d)/blue(e) dots Fig.. 10b as a function of rank of year by PC values (high to low) along with values of PC1 and PC2 for $H_s$ themselves (c and f respectively). Black circles represent values for $95^{th}$ percentile of $H_s$.

**Fig. A1.** Taylor diagram for Hs in reference model simulation (FULL) in comparison to NDBC buoys in DJF 2010.

**Fig. B1.** Spatial distribution of first leading EOFs of seasonal mean storm-track (defined as bandpass-filtered vertically integrated EKE): a, b for synoptic variability of EKE (2-10 days); c, d for low-frequency variability of EKE (>10 days) and e, f for significant wave height.

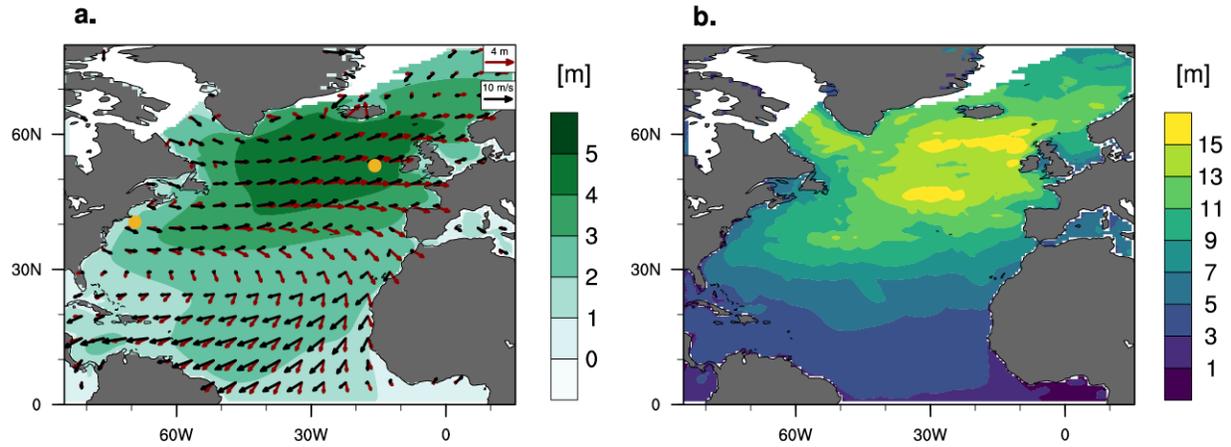

Fig. 1. Simulated climatological seasonal-mean (a) and seasonal-maximum (b) significant wave heights (shading), mean wave and wind directions (red and black vectors respectively) for DJF 1980-2016 (FULL). Colored dots correspond to buoy sites used in further analysis.

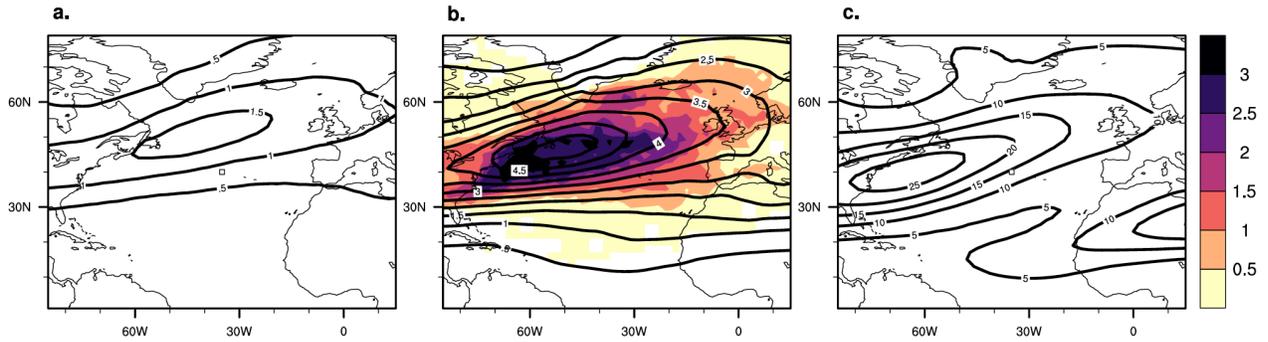

Fig. 2. Climatological seasonal-mean vertically integrated eddy kinetic energy [$\cdot 10^5$ J m$^{-2}$] bandpass-filtered for <2days (a), 2-10 days (b) and >10 days (c) and cyclone track density per season (DJF) per 2-degree box in 1980-2016 (shown in shading on (b)).

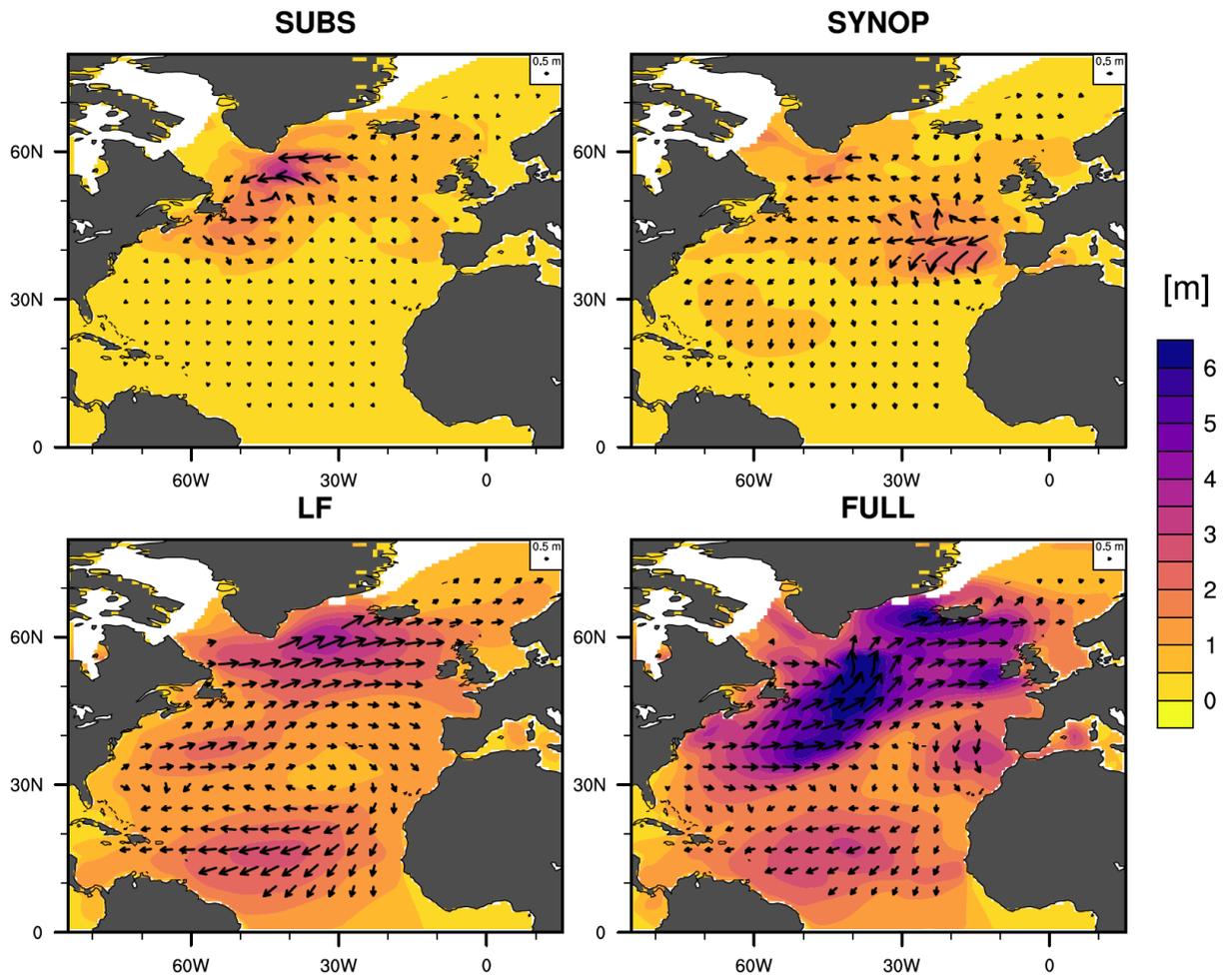

Fig. 3. Snapshots of significant wave heights simulated by the decomposed wind fields (a) SUBS, (b) SYNOP, (c) LF and full forcing (d) FULL at UTC 12:00 12/30/2000.

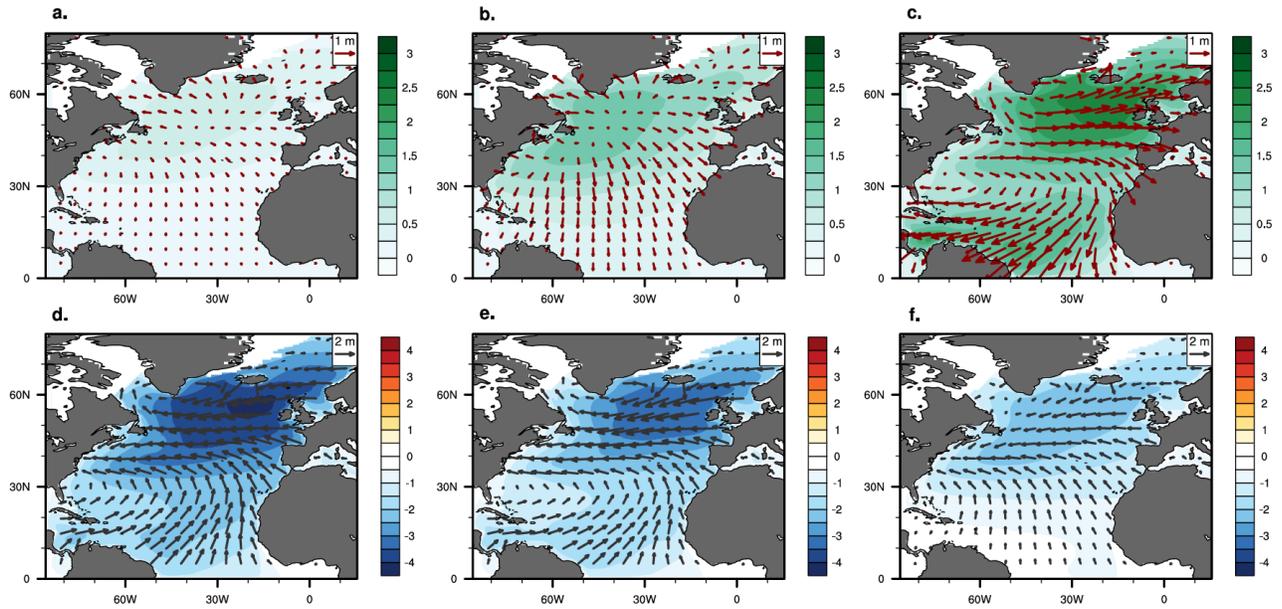

Fig. 4. Climatological seasonal-mean significant wave heights and directions forced by decomposed wind fields: $H_s$ and $\theta$ from SUBS (a), SYNOP (b) and LF (c) simulations and their difference with full forcing simulations: (d) [SUBS-FULL]; (e) [SYNOP-FULL]; (f) [LF-FULL].

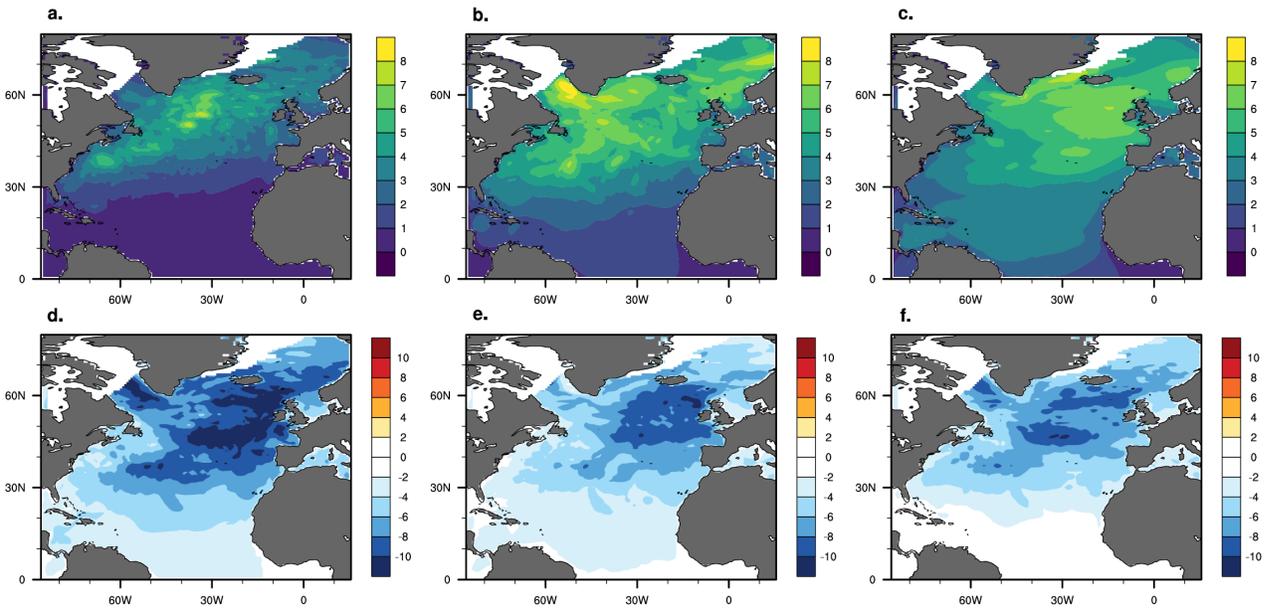

Fig. 5. Maximum significant wave heights forced by decomposed wind fields: $H_s$ from SUBS (a), SYNOP (b) and LF (c) simulations and their difference with full forcing simulations: (d) [SUBS FULL]; (e) [SYNOP-FULL]; (f) [LF-FULL].

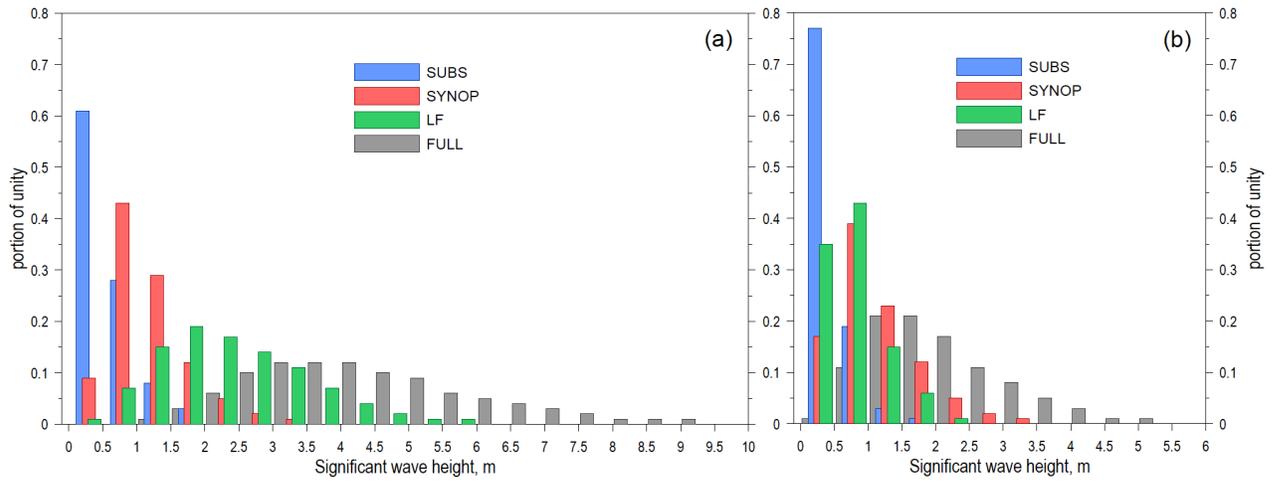

Fig. 6. Histograms of wave heights from SUBS, SYNOP, LF and FULL wave model simulations on the eastern (a) and western (b) margins of the North Atlantic sited in red dots in Fig. 1a.

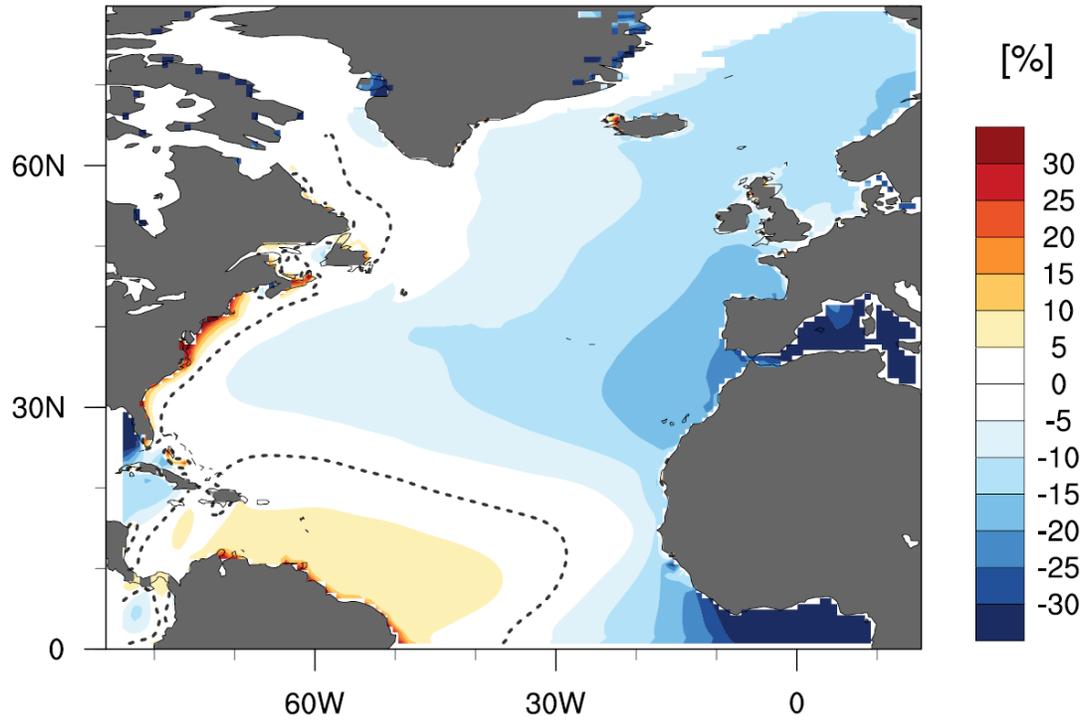

Fig. 7. Difference between sum of wave heights simulated by decomposed forcing and waves simulated by full forcing (i.e. [SUBS + SYNOP + LF] − [FULL]).

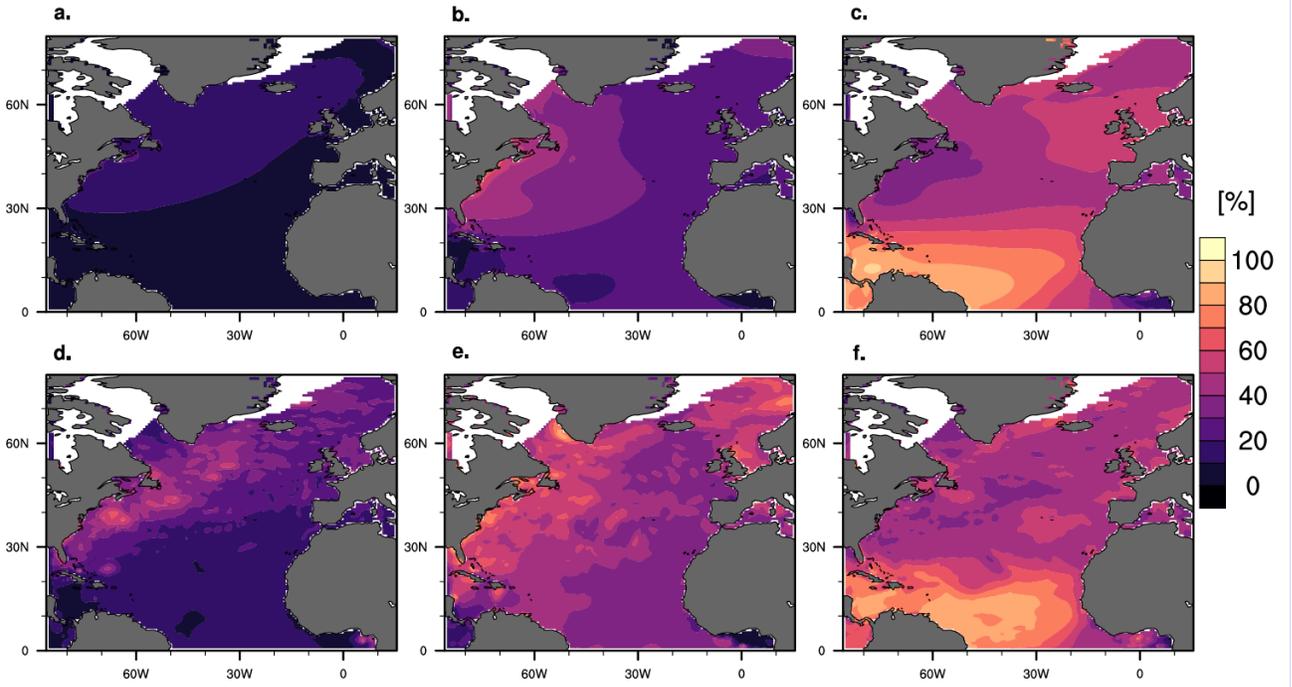

Fig. 8. The ratio between mean (a,b,c) and maximum (d,e,f) significant wave heights from simulations with decomposed forcing and reference simulations: $\frac{[SUBS]}{[FULL]} * 100\%$ (a),(d); $\frac{[SYNOP]}{[FULL]} * 100\%$ (b),(e); $\frac{[LF]}{[FULL]} * 100\%$ (c),(f).

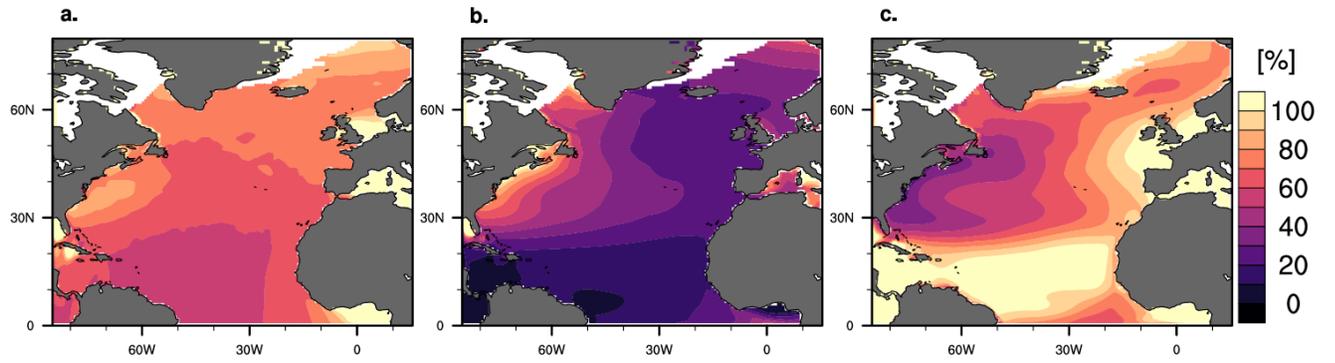

Fig. 9. Ratio between standard deviations of significant wave heights from simulations with decomposed forcing and reference simulations: $\frac{\sigma[SUBS]}{\sigma[FULL]} * 100\%$ (a); $\frac{\sigma[SUBS]}{\sigma[FULL]} * 100\%$ (b); $\frac{\sigma[LF]}{\sigma[FULL]} * 100\%$ (c).

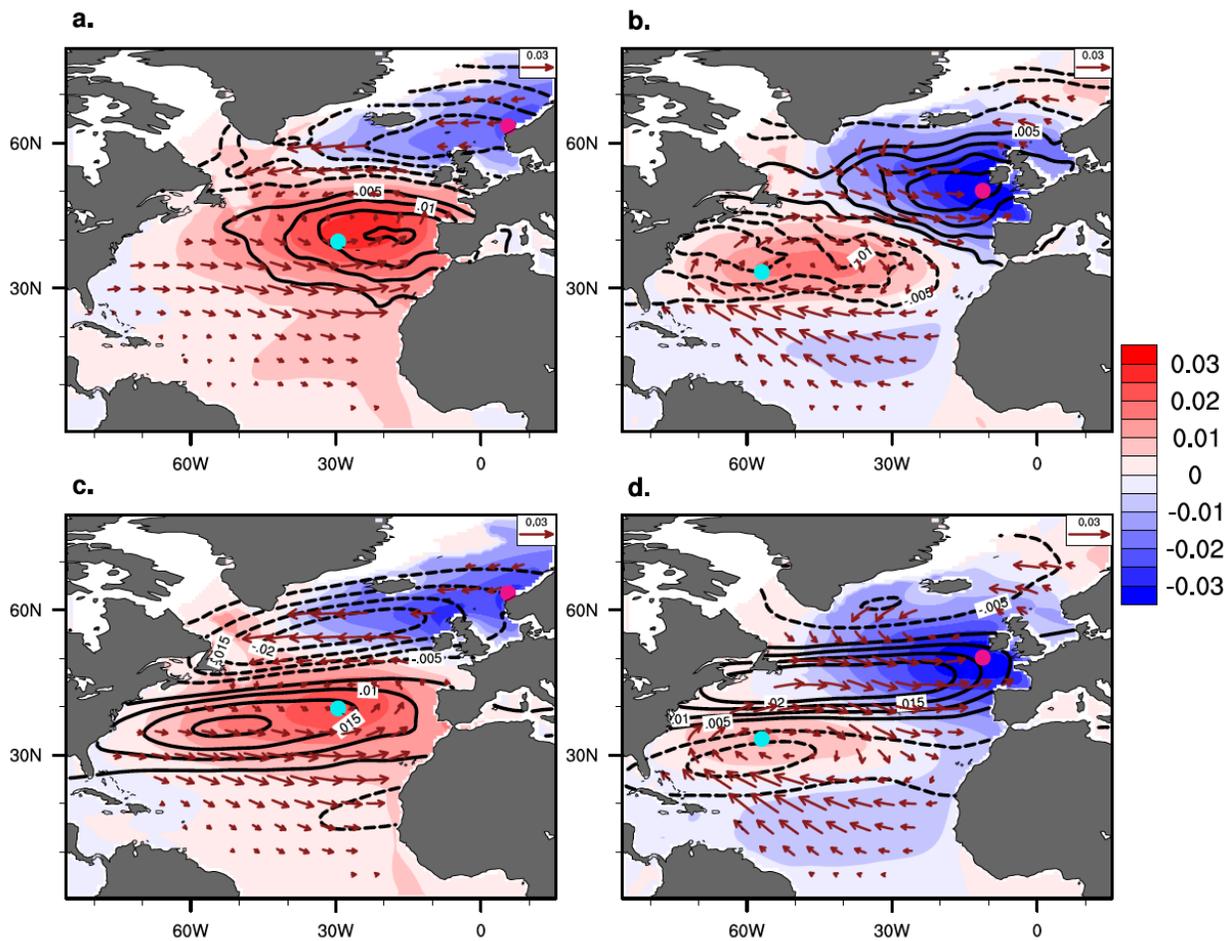

Fig. 10. Pairs for the first two CCA modes between mean $H_s$ and modes of atmospheric variability: 2-10 days band-passed filtered i.e. SYNOP (a,b), and >10 days band-passed filtered i.e. LF, vertically integrated EKE (c,d); a,c represent the first and b,d the second CCA respectively. Solid (dashed) lines indicate positive (negative) values of EKE. Red (blue) scale represents positive (negative) values of the $H_s$ field. Dots are positioned at the locations of maximum (blue) and minimum (purple) values of CCA pattern for $H_s$ and EKE low frequency mode. Vectors indicate CCA modes for wave heights direction ($\theta$).

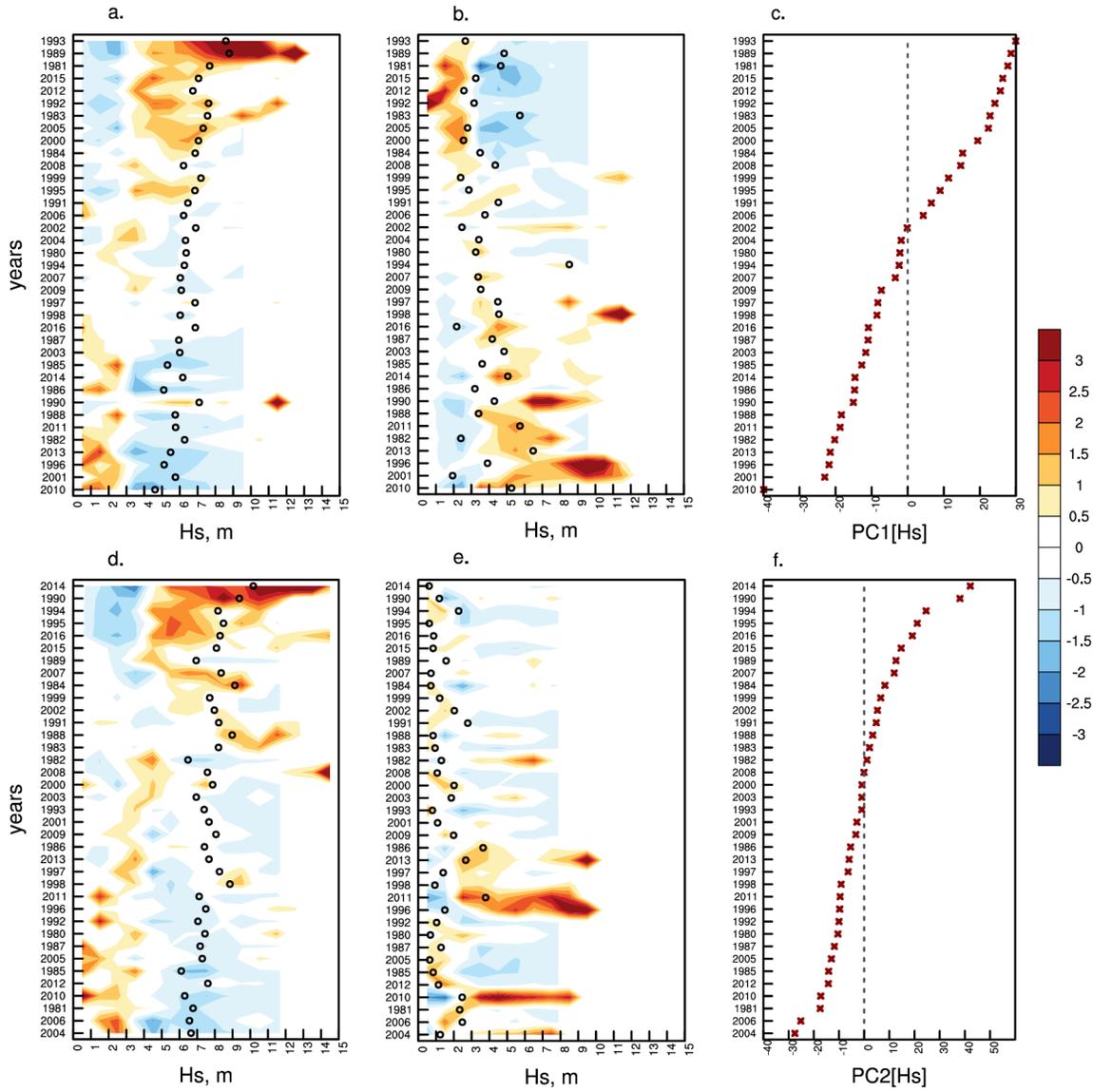

Fig. 11. Normalized occurrence anomalies of $H_s$ in DJF 1980-2016 in the 2-degrees boxes with the centers shown in purple(a)/blue(b) dots in Fig.. 10 a and purple (d)/blue(e) dots Fig.. 10b as a function of rank of year by PC values (high to low) along with values of PC1 and PC2 for $H_s$ themselves (c and f respectively). Black circles represent values for $95^{th}$ percentile of $H_s$.

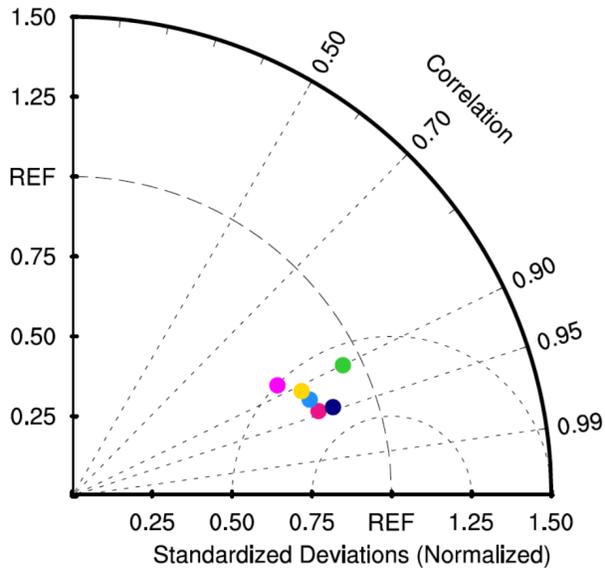 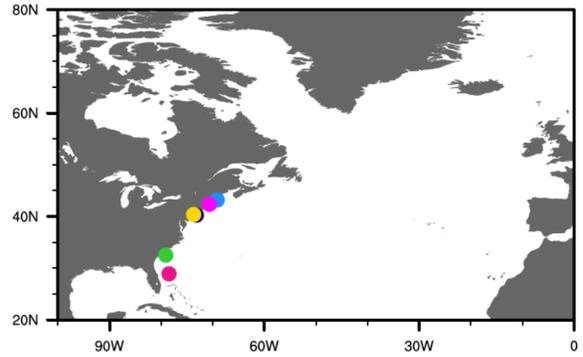

Fig. A1. Taylor diagram for Hs in reference model simulation (FULL) in comparison to NDBC buoys in DJF 2010.

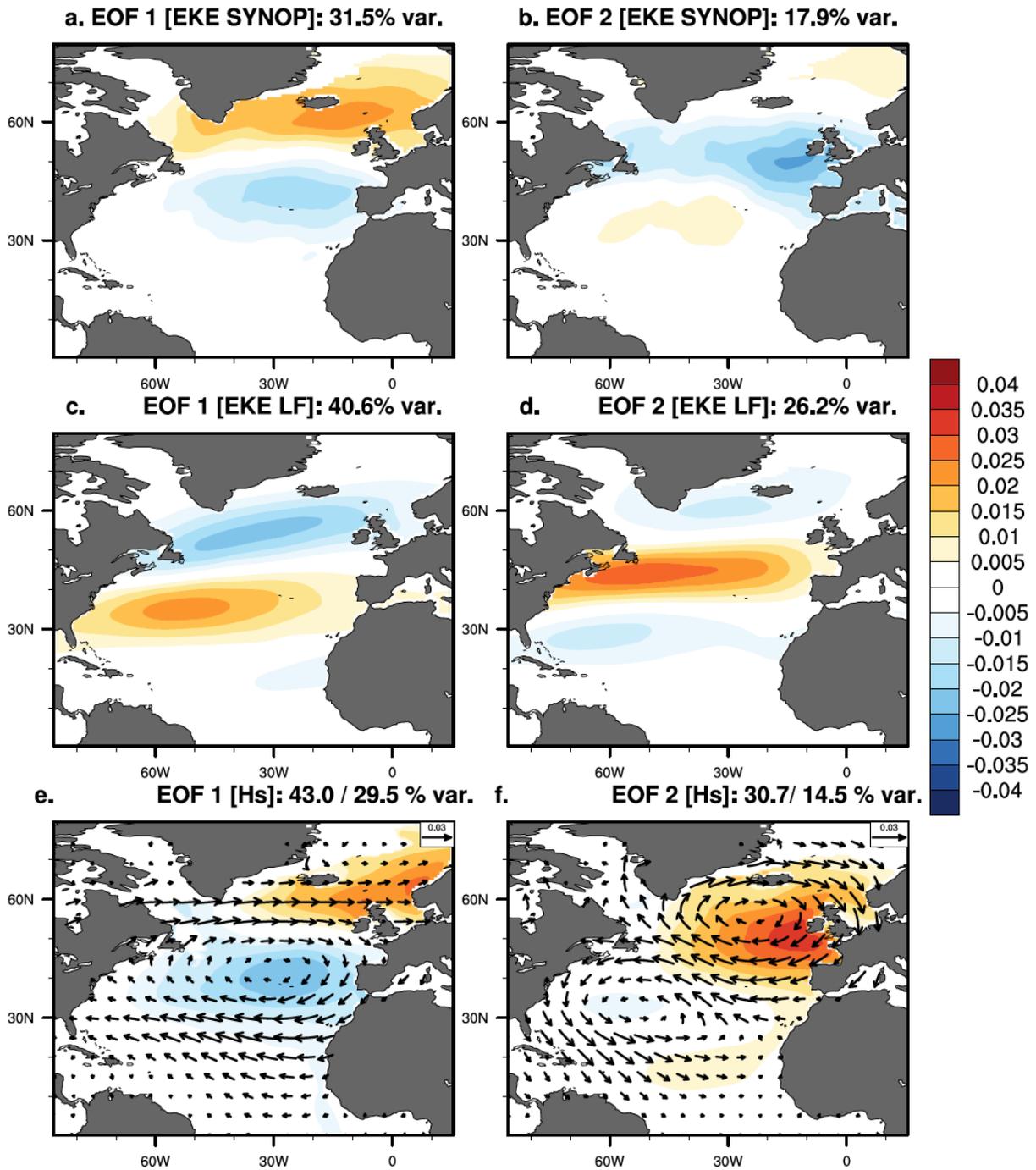

Fig. B1. Spatial distribution of first leading EOFs of seasonal mean storm-track (defined as bandpass-filtered vertically integrated EKE): a, b for synoptic variability of EKE (2-10 days); c, d for low-frequency variability of EKE (>10 days) and e, f for significant wave height and mean wave direction.

56